\documentclass[aps,prl,twocolumn,superscriptaddress,longbibliography,raggedbottom]{revtex4-2}
\usepackage{textcomp}
\usepackage{amsmath}
\usepackage{cancel}
\usepackage{graphicx}
\usepackage{xspace}
\usepackage{amsfonts}
\usepackage{color}

\makeatletter

\definecolor{red}{rgb}{0.75,0,0}
\definecolor{blue}{rgb}{0,0,0.75}
\definecolor{green}{rgb}{0,0.5,0}
\definecolor{orange-red}{RGB}{255,69,0}

\usepackage{dsfont}

\newcommand{\canetset}[1]{{\mathchoice {\hbox{$\sf\textstyle #1\kern-0.4em #1$}}
{\hbox{$\sf\textstyle #1\kern-0.4em #1$}}
{\hbox{$\sf\scriptstyle #1\kern-0.3em #1$}}
{\hbox{$\sf\scriptscriptstyle #1\kern-0.2em #1$}}}}

\def\nbZ{{\mathchoice {\hbox{$\sf\textstyle Z\kern-0.4em Z$}}
{\hbox{$\sf\textstyle Z\kern-0.4em Z$}}
{\hbox{$\sf\scriptstyle Z\kern-0.3em Z$}}
{\hbox{$\sf\scriptscriptstyle Z\kern-0.2em Z$}}}}
\usepackage{bm}

\renewcommand{\exp}[1]{\mathchoice{\mathrm{e}^{#1}}{\operatorname{exp}\left(#1\right)}{\operatorname{exp}\left(#1\right)}{\operatorname{exp}\left(#1\right)}}

\newlength \standardfigwidth
\newcounter{exercise}
{\addtocounter{exercise}{1}\begin{center}\begin{minipage}{0.8\linewidth}\textbf{Exercise
\arabic{exercise}:}\begin{itshape}}% begin code
{\end{itshape}\end{minipage}\end{center}}%                    end code

\makeatletter
\newcommand{\creat}[3][]{\@ifempty{#1}{#2^{\dagger}}{\left(#2^{\dagger}\right)^{#1}}\@ifempty{#3}{}{\!(#3)}}

\newcommand{\creatDoi}[3][]{\@ifempty{#1}{\tilde{#2}}{\left(\tilde{#2}\right)^{#1}}\@ifempty{#3}{}{(#3)}}

\newcommand{\annih}[3][]{#2\@ifempty{#1}{}{^{#1}}\@ifempty{#3}{}{(#3)}}

\makeatother

% ...
\newlength{\bibmarkkeyAleft}

\newlength{\bibmarkkeyBleft}

\newlength{\bibmarkkeyCleft}

\newlength{\bibmarkkeyDleft}

\begin{document}

\title{Stability reversal, activated escape, and mechanical response of negative-mobility particles}

\author{Junteng Li}
%\thanks{Corresponding author: zhyou@xmu.edu.cn}
\affiliation{Fujian Provincial Key Laboratory for Soft Functional Materials Research,
Research Institute for Biomimetics and Soft Matter, Department of
Physics, Xiamen University, Xiamen, Fujian 361005, China}

\author{Zhihong You}
\thanks{Corresponding author: zhyou@xmu.edu.cn}
\affiliation{Fujian Provincial Key Laboratory for Soft Functional Materials Research,
Research Institute for Biomimetics and Soft Matter, Department of
Physics, Xiamen University, Xiamen, Fujian 361005, China}

\date{\today}

%=====================================================
\begin{abstract}
Understanding the fluctuations and response of particles in potential landscapes is a fundamental problem in nonequilibrium statistical physics. Here, we investigate how negative mobility reshapes the stochastic behavior of a particle moving in a mechanical potential landscape. We consider a one-dimensional overdamped particle subject to additive noise, with mobility that is negative at weak forces and positive at strong forces. In a potential well, negative mobility destabilizes the mechanical potential minimum and produces bimodal localization at two off-center positions. The same stability reversal transforms conventional Kramers escape from a potential well into a two-stage escape process involving an intermediate metastable state. When an additional spatially uniform force is applied, the mechanical perturbation translates rather than tilts the effective double well, yielding a positive response dominated at weak noise by intrawell relaxation, even though slow interwell switching dominates spontaneous positional fluctuations. These findings reveal how negative mobility reshapes the interplay among mechanical landscapes, stochastic stability, and response.
\end{abstract}

\maketitle

% ========================================================
\section{Introduction}

How mechanical potentials govern particle fluctuations and response remains a central question in nonequilibrium statistical physics. Specifically, it concerns where particles accumulate in a potential, how long they remain trapped before noise drives them across a barrier, and how they move in response to an additional force. For overdamped Brownian particles at equilibrium, positive linear mobility and the Einstein relation~\cite{Einstein1905} provide a common framework for these phenomena: potential minima define stable states, barriers govern thermally activated escape~\cite{Kramers1940,Hanggi1990}, and fluctuation–dissipation relations connect spontaneous correlations to mechanical response~\cite{Kubo1966,Marconi2008}. Confinement, rare transitions, and response are thus linked through the same underlying dynamics. In nonequilibrium environments, such as active baths~\cite{Marchetti2013,Bechinger2016,Wu2000,Maggi2014,Argun2016} or driven fluids~\cite{Krueger2009}, the relationship between mechanical force, particle drift, and fluctuations need not satisfy these equilibrium constraints~\cite{Harada2005,Blickle2007,Baiesi2009,Seifert2010}. A prescribed potential still determines the applied force, but the resulting stability and escape pathways depend on how the particle moves in response to that force~\cite{Geiseler2016,Caprini2019,Woillez2019,Militaru2021}. Moreover, predicting mechanical response requires specifying how a physical perturbation modifies the dynamics, rather than merely characterizing the unperturbed motion~\cite{Baiesi2013,Dembo2010}. The central challenge is therefore to understand how mechanical potential landscapes and nonequilibrium mechanical response jointly determine the static and dynamic behavior of particles.

Negative mobility, in which particles move against an applied force, represents a striking manifestation of nonequilibrium transport~\cite{Hanggi2009}. Early theoretical studies demonstrated that absolute negative mobility can emerge from interactions between Brownian particles and from the interplay of periodic potentials, nonequilibrium driving, inertia, and thermal fluctuations~\cite{Cleuren2001,Eichhorn2002,Eichhorn2002PRE,Machura2007}. This counterintuitive phenomenon was subsequently demonstrated experimentally in microfluidic systems and driven Josephson junctions~\cite{Ros2005,Nagel2008}. Further studies revealed negative mobility in a variety of complex environments, including inertial tracers in steady fluid flows~\cite{Sarracino2016}, colloidal probes in active nematic fluids~\cite{Foffano2012}, self-propelled Janus particles in corrugated channels~\cite{Ghosh2014}, and active particles advected by fluid flows~\cite{Cecconi2018}. More recently, a microscopic theory demonstrated how interactions with mobile passive crowders can induce absolute negative mobility of active tracers~\cite{Rizkallah2023}. Related studies have also identified negative differential mobility arising from crowding and obstacle-induced trapping, although this does not necessarily imply motion against the applied force~\cite{Zia2002,Leitmann2013,Benichou2014,Reichhardt2018}. Together, these studies establish that negative mobility can arise from diverse nonequilibrium mechanisms. Nevertheless, previous research has primarily focused on particle transport under spatially uniform external driving~\cite{Slapik2019}. In a potential landscape, however, the mechanical force varies with position, allowing particles to move against the force in some regions and along it in others. This competition can reverse the stability of mechanical potential extrema, fundamentally altering particle localization and activated escape. How negative mobility jointly governs localization, activated escape, and mechanical response in potential landscapes therefore remains an important open question.

To address this question, we consider a minimal model of a one-dimensional overdamped particle subject to additive noise, with a force–velocity relation that produces negative mobility at weak forces and positive mobility at strong forces. We show that negative mobility can reverse the conventional stability of mechanical potential extrema: forces that normally restore particles toward potential minima instead drive them away, whereas barrier tops can become attractive rather than repulsive. This stability reversal provides a common physical mechanism for unconventional localization and activated escape in different potential landscapes. In a potential well, the mechanical minimum becomes unstable and two off-center attractors emerge, giving rise to a bimodal stationary distribution that we obtain exactly. In a well with an escape barrier on one side, the same stability reversal transforms the mechanical barrier top into an intermediate metastable state, leading to a two-stage activated escape process for which we derive the exact mean first-passage time. Furthermore, negative mobility produces an unconventional response to external mechanical forcing. Specifically, an additional uniform mechanical force translates rather than tilts the effective double-well landscape, preserving its shape and the relative populations of the two wells. Because both accumulation regions shift together, the mechanical response remains positive and, at weak noise, develops predominantly through rapid intrawell relaxation, even though slow interwell switching dominates spontaneous positional fluctuations. Together, these findings reveal how negative mobility fundamentally reshapes the interplay between mechanical confinement, spontaneous fluctuations, and response to external forces.

% =====================================================
\section{Model}

We consider a one-dimensional overdamped particle moving in a mechanical potential $U(x)$ under stochastic fluctuations. Its position evolves according to the Langevin equation
\begin{equation}
\frac{\mathrm{d}x}{\mathrm{d}t}
=v[f(x)]+\sqrt{2D}\,\eta(t),
\label{eq:langevin}
\end{equation}
where $f(x)=-U'(x)$ is the local mechanical force and $v(f)$ describes the deterministic velocity induced by this force. Here, $D>0$ is the diffusion coefficient, and $\eta(t)$ is a zero-mean Gaussian white noise satisfying $\langle\eta(t)\eta(t')\rangle=\delta(t-t')$. Unlike conventional Brownian particles with constant positive mobility, we allow the deterministic velocity to depend nonlinearly on the applied force, including regimes in which the particle moves against it.

\begin{figure}[!t]
    \centering
    \includegraphics[width=\columnwidth]{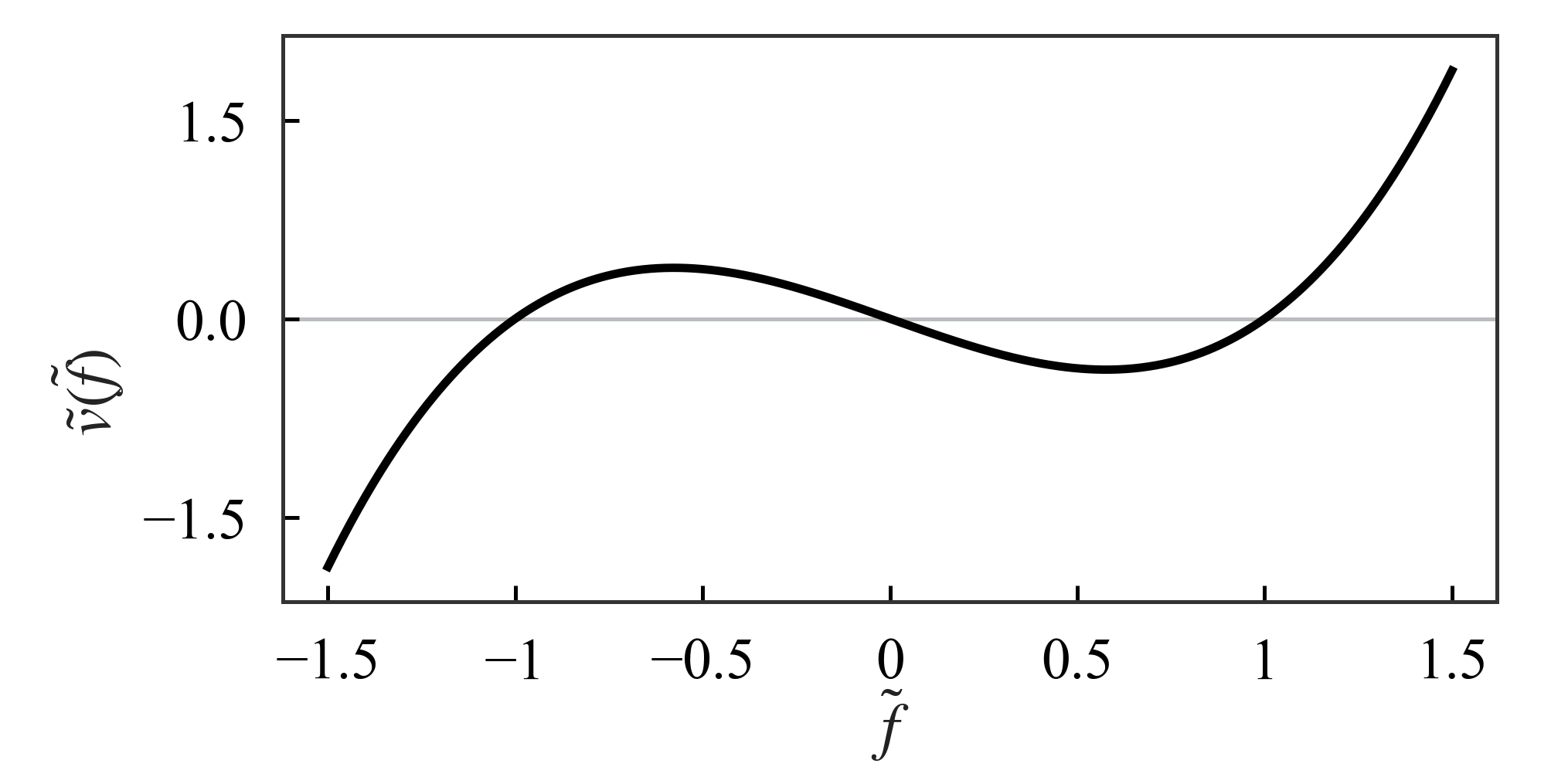}
    \caption{Dimensionless force--velocity relation
    $\tilde v(\tilde f)=\tilde f^3-\tilde f$, corresponding to Eq.~\eqref{eq:vf}.
    The mobility is negative for $0<|\tilde f|<1$ and positive
    for $|\tilde f|>1$, with drift reversal at $\tilde f=\pm1$.}
    \label{fig:01}
\end{figure}

To isolate the consequences of negative mobility, we prescribe the instantaneous force--velocity relation
\begin{equation}
v(f)=v_0\left[\left(\frac{f}{f_0}\right)^3-\frac{f}{f_0}\right],
\label{eq:vf}
\end{equation}
where $v_0>0$ sets the velocity scale and $f_0>0$ is the characteristic force at which the drift reverses direction. This cubic law is the simplest odd polynomial that combines a negative linear response at weak forces with a positive response at sufficiently strong forces. Specifically, the mobility $v(f)/f$ is negative for $0<|f|<f_0$ and positive for $|f|>f_0$ [Fig.~\ref{fig:01}]. Thus, weak forces drive the particle against the force, whereas sufficiently strong forces restore motion along it. The odd symmetry of $v(f)$ ensures that the dynamics has no preferred direction, while $v(0)=0$ excludes deterministic propulsion in the absence of a force.

The model provides a minimal phenomenological description of negative-mobility particles without explicitly resolving the nonequilibrium mechanism sustaining motion against the mechanical force. The instantaneous force--velocity relation neglects memory effects, while the additive noise assumes a spatially and force-independent diffusion coefficient. We treat $D$ as an independent parameter and impose no equilibrium Einstein relation between diffusion and mobility. These assumptions allow us to isolate how negative mobility modifies particle localization, activated escape, and mechanical response in potential landscapes, while leaving its microscopic origin and associated energetic cost outside the scope of the present study. The probability density $p(x,t)$ associated with Eq.~\eqref{eq:langevin} obeys the Fokker--Planck equation~\cite{Risken1989},
\begin{equation}
\frac{\partial p}{\partial t}
=-\frac{\partial}{\partial x}\left[v[f(x)]p\right]
+D\frac{\partial^2p}{\partial x^2}
=-\frac{\partial J}{\partial x},
\label{eq:fpe}
\end{equation}
where $J(x,t)=v[f(x)]p(x,t)-D\,\partial_xp(x,t)$ is the probability current. The drift contribution transports probability along the deterministic velocity, which can oppose the local mechanical force, whereas the diffusive contribution acts to smooth spatial variations in the density. 

We nondimensionalize the model using the characteristic length $\ell_0=D/v_0$, time $\tau_0=D/v_0^2$, and energy $E_0=f_0D/v_0$, with the corresponding force unit $f_0$. The length and time scales compare deterministic motion at speed $v_0$ with diffusive spreading, while $E_0=f_0\ell_0$ is a mechanical work scale, not an assumed thermal energy. Hereafter, dimensionless quantities are denoted by a tilde, while intrinsically dimensionless parameters are written without one. The dimensionless Langevin equation is integrated using the Euler--Maruyama scheme~\cite{Kloeden1992} with time step $\delta\tilde t=10^{-3}$ ($2\times10^{-5}$ for the strong-noise data in Fig.~\ref{fig:02}(d)). Stationary ensembles are prepared by releasing particles at $\tilde x=0$, which by symmetry populates both wells equally, and relaxing them without additional force for a time much longer than the intrawell relaxation time. In the response measurements, trajectories under $\pm\tilde F$ start from the same stationary configurations and share the same noise realizations, so that statistical fluctuations largely cancel in their difference. Each parameter set is sampled with more than $5\times 10^5$ independent trajectories, and error bars, given by standard errors over independent runs, are smaller than the symbols.

% =====================================================

\section{Statistical distribution in a potential well}

\begin{figure*}[t]
    \centering
    \includegraphics[width=\textwidth]{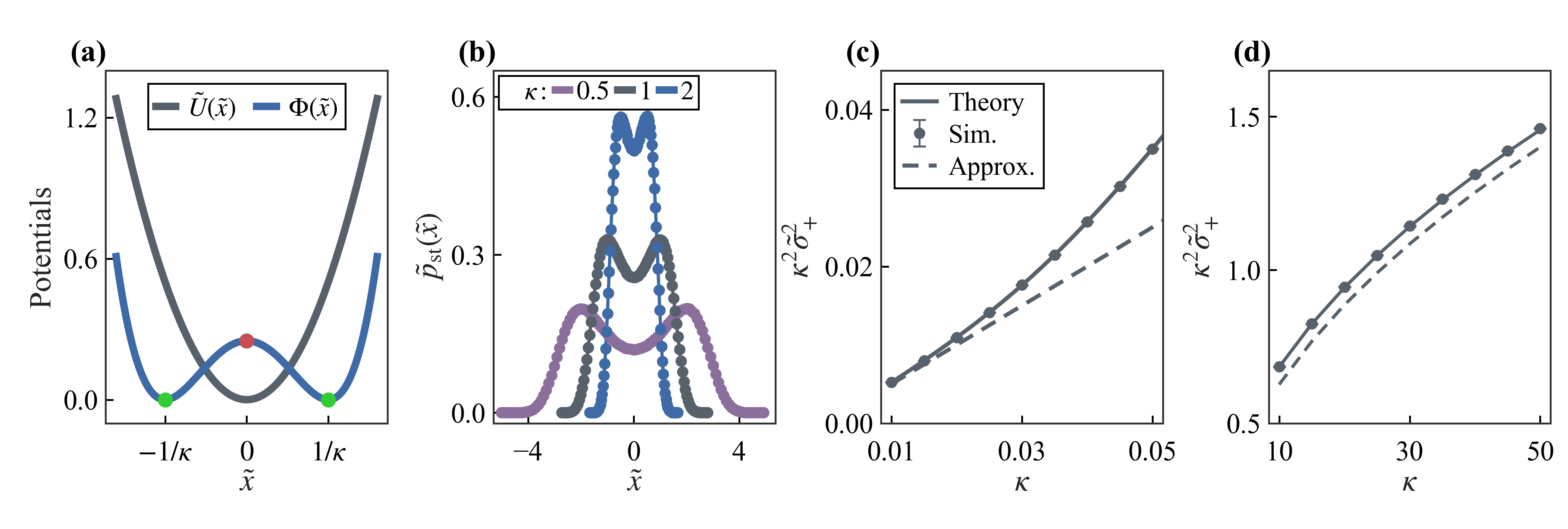}
\caption{Stationary localization in a harmonic trap. (a) $\tilde U(\tilde x)$ and $\Phi(\tilde x)$ for $\kappa=1$; green (red) points mark stable (unstable) fixed points of $\tilde B_0$. (b) Stationary densities for several $\kappa$. (c), (d) Normalized conditional variance $\kappa^2\tilde\sigma_+^2$ at small (c) and large (d) $\kappa$, with the asymptotic predictions $\kappa/2$ [Eq.~\eqref{eq:sigmap}] and $0.198\sqrt{\kappa}$ [Eq.~(S8)] (dashed). Lines are exact results and points are simulations.}
    \label{fig:02}
\end{figure*}

We first investigate confinement of a negative-mobility particle in a mechanical potential well. For simplicity, we consider a harmonic potential $\tilde U(\tilde x)=\kappa\tilde x^2/2$, where $\kappa=kD/(v_0f_0)$ is the dimensionless stiffness and $k$ the dimensional one [Fig.~\ref{fig:02}(a)]. The restoring force $\tilde f=-\kappa\tilde x$ generates the drift $\tilde B_0(\tilde x)=\kappa\tilde x-\kappa^3\tilde x^3$. Because the dimensionless diffusion coefficient is unity, $\kappa$ is the only control parameter. Numerical simulations reveal a symmetric, bimodal stationary distribution, where particles are depleted from the mechanical potential minimum and accumulate near two off-center positions, $\tilde x_\pm=\pm\kappa^{-1}$ [Fig.~\ref{fig:02}(b)]. This off-center accumulation contrasts with ordinary harmonic confinement, in which particles preferentially occupy the mechanical potential minimum. The accumulation positions are selected by the reversal of the force--velocity relation. Within $|\tilde x|<\kappa^{-1}$, the restoring force lies on the negative-mobility branch and drives particles away from the trap center. Beyond these positions, positive mobility restores inward motion. Consequently, the mechanical minimum becomes unstable, while stable attractors appear where $|\tilde f|=1$, at $\tilde x_\pm=\pm1/\kappa$ [Fig.~\ref{fig:02}(a)]. We refer to the destabilization of mechanical minima and stabilization of mechanical maxima under negative mobility as stability reversal. In the harmonic trap, the recovery of positive mobility at stronger forces additionally creates the two off-center attractors.

The stationary density follows from the zero-current condition of the Fokker--Planck equation, $\tilde J_{\mathrm{st}}=\tilde B_0\tilde p_{\mathrm{st}}-\partial_{\tilde x}\tilde p_{\mathrm{st}}=0$. Its solution is $\tilde p_{\mathrm{st}}(\tilde x)=\tilde Z^{-1}e^{-\Phi(\tilde x)}$, where $\tilde Z$ is a normalizing constant and $\tilde B_0=-\Phi'$ defines the effective drift potential~\cite{Risken1989}. Substituting $\tilde f=-\tilde U'$ into the force--velocity relation gives the effective drift potential $\Phi(\tilde x)=\int^{\tilde x}[\tilde U'(y)]^3\,\mathrm{d}y-\tilde U(\tilde x)$, up to an additive constant. For the harmonic trap, choosing $\Phi(\tilde x_\pm)=0$ yields $\Phi(\tilde x)=(\kappa^2\tilde x^2-1)^2/(4\kappa)$. The nonlinear force--velocity relation therefore converts a single-well mechanical potential into an effective double well [Fig.~\ref{fig:02}(a)], with the exact stationary density
\begin{equation}
  \tilde{p}_{\mathrm{st}}(\tilde{x})
  = \frac{1}{\tilde{Z}}
    \exp {-\frac{(\kappa^{2}\tilde{x}^{2}-1)^{2}}{4\kappa}},
  \label{eq:pst}
\end{equation}
agreeing well with those from the simulations [Fig.~\ref{fig:02}(b)].
The density maxima coincide with the deterministic attractors $\tilde x_\pm=\pm\kappa^{-1}$ for every finite $\kappa>0$. The effective barrier height, $\Delta\Phi=\Phi(0)-\Phi(\tilde x_+)=1/(4\kappa)$, determines the peak-to-center ratio, $\tilde p_{\mathrm{st}}(\tilde x_\pm)/\tilde p_{\mathrm{st}}(0)= \exp{1/(4\kappa)}$. Increasing $\kappa$ therefore weakens the central depletion. At fixed dimensional trap stiffness and force--velocity parameters, this corresponds to increasing the dimensional noise strength. Nevertheless, the origin remains a local minimum of the stationary density for every finite $\kappa>0$. Noise thus fills the central depletion continuously, without inducing a transition to a unimodal distribution, even when the two peaks become difficult to resolve.

To quantify the width of a single accumulation region without including the separation between the populations, we use the conditional variance $\tilde\sigma_+^2=\langle\tilde x^2\rangle_+-\langle\tilde x\rangle_+^2$, where $\langle\cdot\rangle_+$ denotes the stationary average $\langle\cdot\rangle_{\mathrm{st}}$ restricted to $\tilde x>0$. In the weak-noise regime, $\Delta\Phi\gg1$, or equivalently $4\kappa\ll1$, particles in the right population remain close to $\tilde x_+=1/\kappa$. Small deviations relax at the rate $-\tilde B_0'(\tilde x_+)=\Phi''(\tilde x_+)=2\kappa$. Since $\tilde B_0'(\tilde x)=-\kappa\tilde v'(-\kappa\tilde x)$, this rate is the product of the trap stiffness and the differential mobility at the reversal force, $\tilde v'(-1)=2$. Although the mobility $\tilde v(\tilde f)/\tilde f$ vanishes at $|\tilde f|=1$, the slope of the force--velocity curve remains positive and provides local confinement. Balancing diffusion against this restoring drift gives an approximately Gaussian population with 
\begin{equation}
    \label{eq:sigmap}
    \tilde\sigma_+^2\simeq\frac{1}{\Phi''(\tilde x_+)} =\frac{1}{2\kappa}.
\end{equation}
This yields a normalized variance $\kappa^2\tilde\sigma_+^2 \simeq \kappa/2$ measuring the population width relative to the squared distance of the attractor from the trap center. It therefore separates physical broadening from the change in the noise-dependent length unit and is the quantity shown in Fig.~\ref{fig:02}(c).

The leading corrections to the Gaussian approximation reflect the asymmetry of each effective well. Near the right attractor, the restoring drift is weaker toward the central barrier and stronger toward the outer flank than its linearization predicts. The softer inner flank broadens the population beyond the local Gaussian estimate [Fig.~\ref{fig:02}(c)] and shifts its conditional mean toward the trap center. In the small-stiffness regime, $\kappa\ll1$, the effective barrier $\Delta\Phi=1/(4\kappa)$ is large relative to the fixed diffusion scale, and $\kappa\langle\tilde x\rangle_+ =1-\frac{3}{4}\kappa+O(\kappa^2)$ (see Eq.~(S6) in the Supplemental Material~\cite{SM}). The conditional mean therefore lies closer to the trap center than the density maximum at $\tilde x_+=1/\kappa$. By symmetry, $\langle\tilde x\rangle_{\mathrm{st}}=0$ and $\langle\tilde x^2\rangle_{\mathrm{st}} =\tilde\sigma_+^2+\langle\tilde x\rangle_+^2$, giving $\kappa^2\langle\tilde x^2\rangle_{\mathrm{st}} =1-\kappa+O(\kappa^2)$.
Here, $\kappa^2\langle\tilde x^2\rangle_{\mathrm{st}}$ is the full variance normalized by the squared force-reversal distance, $\tilde x_+^2=\kappa^{-2}$. Its decrease shows that the inward shift of the conditional means outweighs the broadening of the individual populations, producing an additional contraction relative to the attractor separation. The leading behavior $\langle\tilde x^2\rangle_{\mathrm{st}}\sim\kappa^{-2}$ reflects the separation of the two populations, which decreases as the trap stiffens. This differs from ordinary harmonic confinement, where the variance scales as $\kappa^{-1}$ at fixed noise strength and positive mobility.

For large stiffness, $\kappa\gg1$, the distribution narrows in absolute units of $\ell_0$ but extends increasingly far beyond the attractors relative to their separation. The quartic term dominates the effective drift potential, $\Phi(\tilde x)\simeq\kappa^3\tilde x^4/4$, setting the typical scale $|\tilde x|\sim\kappa^{-3/4}$. The corresponding force magnitude, $|\tilde f|=\kappa|\tilde x|\sim\kappa^{1/4}$, lies predominantly on the positive-mobility branch, where $\tilde v(\tilde f)\simeq\tilde f^3$. Consequently,
\[
\tilde\sigma_+^2\simeq0.198\,\kappa^{-3/2},
\qquad
\kappa^2\tilde\sigma_+^2\simeq0.198\sqrt{\kappa}
\]
[Eq.~(S8) and Fig.~\ref{fig:02}(d)].
Thus, the conditional variance decreases with stiffness more rapidly than the $\kappa^{-1}$ scaling of ordinary harmonic confinement, reflecting the stronger cubic restoring drift. Nevertheless, its value relative to the squared force-reversal distance increases because the attractors approach the trap center faster than the distribution narrows.

These results identify stability reversal and force-selected attraction, rather than mechanical energy minimization, as the organizing mechanism of confinement in this model. Off-center accumulation in harmonic traps is also known for self-propelled particles, where persistence and the balance between propulsion and the restoring force can produce localization away from the trap center~\cite{Tailleur2009,Pototsky2012,Takatori2016,Dhar2019,Malakar2020}. Here, the accumulation instead follows from the local force--velocity relation, without requiring persistent self-propulsion or a mechanically bistable potential. Although the stationary density has a Boltzmann-like form in $\Phi$, this representation follows from the one-dimensional zero-current condition and does not imply equilibrium with respect to the mechanical potential $U$. Because the effective drift potential rather than the mechanical potential determines stochastic stability, negative mobility also reorganizes activated escape, as considered next.

% =====================================================
\section{Kramers escape}
\label{sec:escape}

\begin{figure*}
    \centering
    \includegraphics[width=\textwidth]
    {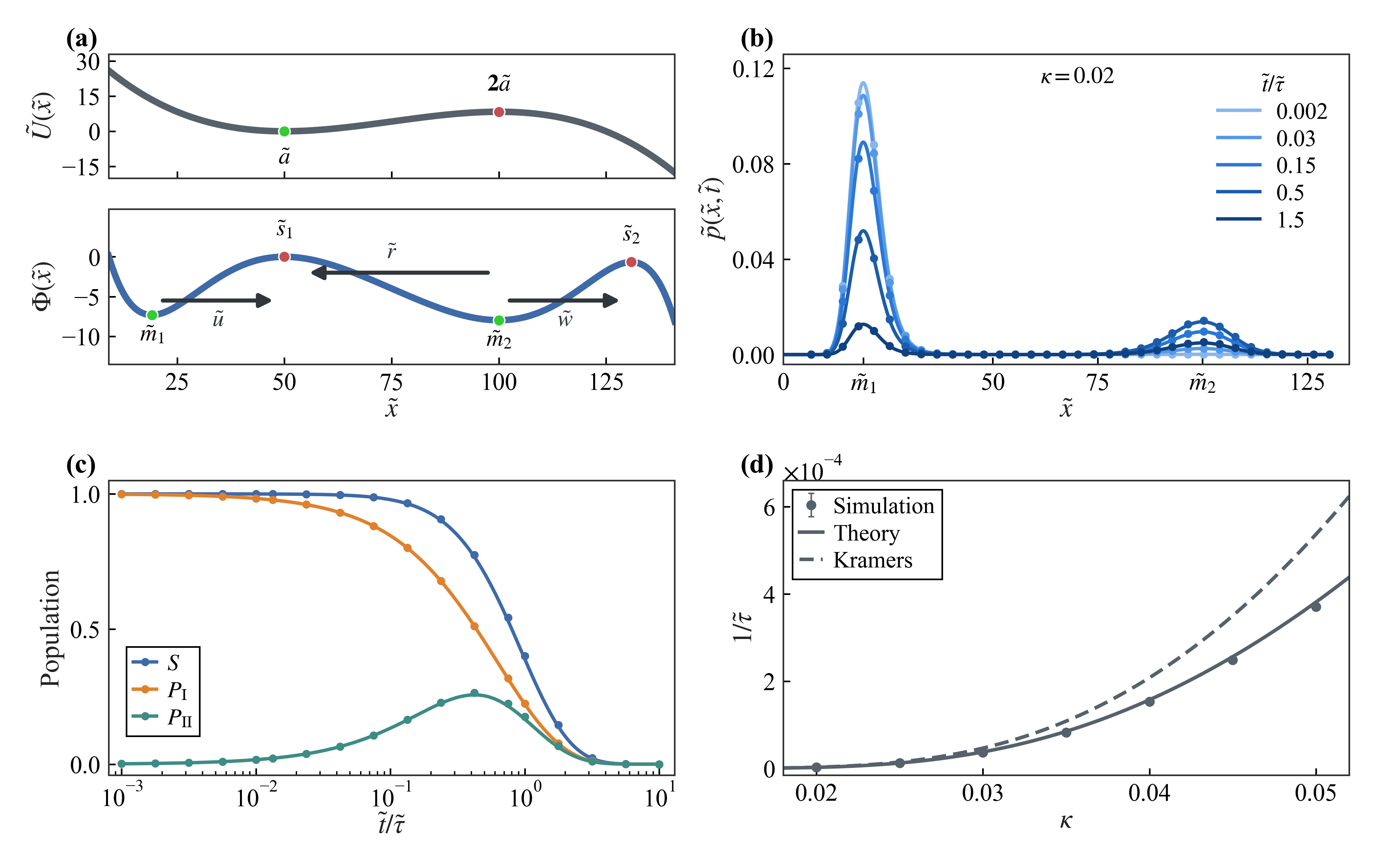}
    \caption{First passage in the cubic trap, Eq.~\eqref{eq:cubicU}, at $\kappa \tilde a=1$. (a) $\tilde U(\tilde x)$ (top) and $\Phi(\tilde x)$ (bottom) for $\kappa=0.02$; green (red) points mark the mechanical minimum (maximum) in the top panel and stable (unstable) fixed points of $\tilde B_0$ in the bottom panel. Arrows indicate the rates $\tilde u$, $\tilde r$, and $\tilde w$. (b) Density $\tilde p(\tilde x,\tilde t)$ and (c) probability $P_{\mathrm{I}}$ and $P_{\mathrm{II}}$ that the particle is in the initial basin ($\tilde x<\tilde s_1$) or the intermediate basin ($\tilde s_1<\tilde x<\tilde s_2$), with survival probability $S=P_{\mathrm{I}}+P_{\mathrm{II}}$, for $\kappa=0.02$. (d) Inverse MFPT versus $\kappa$, with the high-barrier (Kramers) approximation $1/(\tilde\tau_{11}+\tilde\tau_{21}+\tilde\tau_{22})$ from Eq.~\eqref{eq:mfpt-integral-asymptotics} (dashed). Lines are Fokker--Planck results [Eq.~\eqref{eq:mfpt-exact-integral} in (d)] and points are simulations.}
    \label{fig:03}
\end{figure*}

Negative mobility also profoundly changes the escape process from a potential barrier. To show this, we consider the cubic mechanical potential
\begin{equation}
\tilde U(\tilde x)=\frac{\kappa}{2}(\tilde x-\tilde a)^2-\frac{\kappa}{3\tilde a}(\tilde x-\tilde a)^3,
\label{eq:cubicU}
\end{equation}
where $\kappa=kD/(v_0f_0)$ is the dimensionless curvature at the mechanical minimum, with $k=U''(a)$, and $\tilde a=a/\ell_0>0$ is the separation between the mechanical minimum at $\tilde x=\tilde a$ and the barrier top at $\tilde x=2\tilde a$. The corresponding force is $\tilde f(\tilde x)=(\kappa/\tilde a)(\tilde x-\tilde a)(\tilde x-2\tilde a)$, and $\kappa\tilde a=ka/f_0$ measures its characteristic magnitude relative to the force-reversal threshold. The four-fixed-point structure described below holds for $0<\kappa \tilde  a<4$. We retain $\kappa \tilde a=1$, so that $\tilde a=\kappa^{-1}$ and $\kappa$ is the only independent control parameter. Because $\kappa \tilde a$ is independent of $D$, this choice fixes the mechanical potential through $a=f_0/k$, and varying $\kappa$ at fixed $k$ corresponds to varying the noise strength. The force then becomes $\tilde f(\tilde x)=(\kappa\tilde x-1)(\kappa\tilde x-2)$.
Stability reversal acts on both mechanical extrema: the minimum at $\tilde x=\kappa^{-1}$ becomes unstable, whereas the barrier top at $\tilde x=2\kappa^{-1}$ becomes an attractor [Fig.~\ref{fig:03}(a)]. The effective drift potential $\Phi$ therefore has two minima $\tilde m_i$ and two maxima $\tilde s_j$, ordered as $\tilde m_1<\tilde s_1<\tilde m_2<\tilde s_2$, with $\tilde m_1=(3-\sqrt{5})/(2\kappa)$, $\tilde s_1=1/\kappa$, $\tilde m_2=2/\kappa$, and $\tilde s_2=(3+\sqrt{5})/(2\kappa)$. Here, $\tilde s_1$ and $\tilde m_2$ coincide with the mechanical extrema, where $\tilde f=0$, whereas $\tilde m_1$ and $\tilde s_2$ occur at the force-reversal threshold $\tilde f=1$. Thus, the mechanical barrier top, which ordinarily separates a trapped state from escape, becomes an intermediate trapping state.

We investigate the mean first-passage time (MFPT) $\tilde\tau$ of a particle initially located at the stable state $\tilde m_1$, with an absorbing boundary at $\tilde s_2$, where the particle is removed upon its first arrival. The particle first crosses the effective barrier at $\tilde s_1$ into the basin around $\tilde m_2$, from which it can either return to the initial basin or reach $\tilde s_2$. For $\kappa \tilde a=1$, the effective barrier heights scale as $\kappa^{-1}$, and sufficiently weak noise produces a two-stage activated process. Figure~\ref{fig:03}(b) shows this sequence directly: probability transiently accumulates around $\tilde m_2$ before absorption. Accordingly, the probability $P_{\mathrm{II}}$ of finding the particle in the intermediate basin rises to about $0.25$ at $\tilde t\approx0.4\tilde\tau$, while the survival probability $S$ stays near unity at early times because absorption requires passing through the intermediate basin [Fig.~\ref{fig:03}(c)].  

The backward Fokker--Planck equation, with absorption at
$\tilde s_2$ and confinement as $\tilde x\to-\infty$, gives the exact
MFPT~\cite{Gardiner2009,Redner2001,Hanggi1990},
\begin{equation}
    \tilde\tau(\tilde m_1) =
    \int_{\tilde m_1}^{\tilde s_2}
    \mathrm{d}y\,e^{\Phi(y)}
    \int_{-\infty}^{y}
    \mathrm{d}z\,e^{-\Phi(z)}.
    \label{eq:mfpt-exact-integral}
\end{equation}
Thus, escape is directly governed by the effective drift potential $\Phi$ rather than the mechanical potential $U$. Splitting the outer integral at $\tilde m_2$ and the inner integral at $\tilde s_1$ yields the decomposition $\tilde\tau(\tilde m_1)=\tilde\tau_{11}+\tilde\tau_{21}+\tilde\tau_{22}$, where
\begin{subequations}
\label{eq:mfpt-integral-decomposition}
\begin{align}
    \tilde\tau_{11}
    &=
    \int_{\tilde m_1}^{\tilde m_2}
    \mathrm{d}y\,e^{\Phi(y)}
    \int_{-\infty}^{y}
    \mathrm{d}z\,e^{-\Phi(z)},
    \label{eq:I11-int}\\
    \tilde\tau_{21}
    &=
    \left[
    \int_{\tilde m_2}^{\tilde s_2}
    \mathrm{d}y\,e^{\Phi(y)}
    \right]
    \left[
    \int_{-\infty}^{\tilde s_1}
    \mathrm{d}z\,e^{-\Phi(z)}
    \right],
    \label{eq:I21-int}\\
    \tilde\tau_{22}
    &=
    \int_{\tilde m_2}^{\tilde s_2}
    \mathrm{d}y\,e^{\Phi(y)}
    \int_{\tilde s_1}^{y}
    \mathrm{d}z\,e^{-\Phi(z)}.
    \label{eq:I22-int}
\end{align}
\end{subequations}

The first and second indices identify the barrier and basin, respectively, that dominate each integral in the activated regime. For effective barriers large compared with the unit diffusion scale, Laplace expansion~\cite{Bovier2004} yields
\begin{subequations}
\label{eq:mfpt-integral-asymptotics}
\begin{align}
    \tilde\tau_{11}
    &\simeq
    \frac{2\pi}{\sqrt{\Phi''(\tilde m_1)\,|\Phi''(\tilde s_1)|}}
    e^{\Phi(\tilde s_1)-\Phi(\tilde m_1)},
    \label{eq:I11}\\
    \tilde\tau_{21}
    &\simeq
    \frac{\pi}{\sqrt{\Phi''(\tilde m_1)\,|\Phi''(\tilde s_2)|}}
    e^{\Phi(\tilde s_2)-\Phi(\tilde m_1)},
    \label{eq:I21}\\
    \tilde\tau_{22}
    &\simeq
    \frac{\pi}{\sqrt{\Phi''(\tilde m_2)\,|\Phi''(\tilde s_2)|}}
    e^{\Phi(\tilde s_2)-\Phi(\tilde m_2)},
    \label{eq:I22}
\end{align}
\end{subequations}
where $\Phi''(\tilde m_i)>0$ and $\Phi''(\tilde s_j)<0$ are the curvatures of the effective drift potential at its minima and maxima.
The factors $\pi$, rather than $2\pi$, in the last two terms arise because absorption at $\tilde s_2$ retains only half
of the Gaussian contribution around that barrier.

To clarify the physical meaning of the three contributions, we consider a two-basin kinetic description in the high-barrier regime, where intrawell relaxation is much faster than activated transitions. Let $\tilde u$ denote the transition rate from the initial basin around $\tilde m_1$ to the intermediate basin around $\tilde m_2$, $\tilde r$ the reverse transition rate, and $\tilde w$ the rate of absorption at $\tilde s_2$ from the intermediate basin. The corresponding Kramers approximations are~\cite{Kramers1940,Hanggi1990}
\begin{subequations}
\label{eq:mfpt-rates}
\begin{align}
    \tilde u
    &\simeq
    \frac{\sqrt{\Phi''(\tilde m_1)\,|\Phi''(\tilde s_1)|}}{2\pi}
    e^{-[\Phi(\tilde s_1)-\Phi(\tilde m_1)]},
    \label{eq:rate-u}\\
    \tilde r
    &\simeq
    \frac{\sqrt{\Phi''(\tilde m_2)\,|\Phi''(\tilde s_1)|}}{2\pi}
    e^{-[\Phi(\tilde s_1)-\Phi(\tilde m_2)]},
    \label{eq:rate-r}\\
    \tilde w
    &\simeq
    \frac{\sqrt{\Phi''(\tilde m_2)\,|\Phi''(\tilde s_2)|}}{\pi}
    e^{-[\Phi(\tilde s_2)-\Phi(\tilde m_2)]}.
    \label{eq:rate-w}
\end{align}
\end{subequations}
The prefactor of $\tilde w$ is twice the conventional Kramers rate because absorption occurs upon first reaching the barrier top $\tilde s_2$, rather than after crossing it.

Each visit to the intermediate basin ends in one of two ways: the particle exits to the left and returns to $\tilde m_1$ with probability $P_L=\tilde r/(\tilde r+\tilde w)$, or it reaches $\tilde s_2$ and is absorbed with probability $1-P_L=\tilde w/(\tilde r+\tilde w)$. The mean duration of each visit is $1/(\tilde r+\tilde w)$, and the number of returns follows a geometric distribution with mean $\tilde r/\tilde w$. Consequently, each basin is visited $1+\tilde r/\tilde w$ times on average. Since each residence in the initial basin lasts $1/\tilde u$, the total MFPT becomes
\begin{equation}
\begin{aligned}
    \tilde\tau(\tilde m_1)
    &\simeq
    \left(1+\frac{\tilde r}{\tilde w}\right)
    \left(\frac{1}{\tilde u}+\frac{1}{\tilde r+\tilde w}\right)\\
    &=\frac{1}{\tilde u}+\frac{\tilde r}{\tilde u\tilde w}+\frac{1}{\tilde w}.
\end{aligned}
\label{eq:mfpt-three-contributions}
\end{equation}
Comparing this expression with Eq.~\eqref{eq:mfpt-integral-asymptotics} establishes the correspondence
\begin{equation}
    \tilde\tau_{11}\simeq\frac{1}{\tilde u},
    \qquad
    \tilde\tau_{21}\simeq\frac{\tilde r}{\tilde u\tilde w},
    \qquad
    \tilde\tau_{22}\simeq\frac{1}{\tilde w}.
    \label{eq:mfpt-integral-process-correspondence}
\end{equation}
Thus, $\tilde\tau_{11}$ describes the initial residence in the left basin, $\tilde\tau_{21}$ the additional residence there caused by returns, and $\tilde\tau_{22}$ the cumulative residence in the intermediate basin over all visits. While the integral decomposition is exact, this kinetic interpretation applies in the high-barrier regime.

The analytical results reveal that negative mobility fundamentally changes the factors determining the escape time. In conventional Kramers escape, the MFPT is primarily controlled by the barrier separating a mechanical potential minimum from its escape state. Here, stability reversal creates two effective barriers and an intermediate metastable state, so the MFPT depends on the residence times in both basins and the possibility of returns between them. As shown by Eq.~\eqref{eq:mfpt-three-contributions}, the two activated passages contribute $1/\tilde u$ and $1/\tilde w$, while returns introduce an additional contribution $\tilde r/(\tilde u\tilde w)$. For the chosen cubic potential with $\kappa \tilde a=1$, the two effective barriers have equal heights, $\Phi(\tilde s_1)-\Phi(\tilde m_1)=\Phi(\tilde s_2)-\Phi(\tilde m_2)\propto\kappa^{-1}$, so both activated stages contribute at the same leading exponential order, with $\tilde\tau_{11}\simeq 2\tilde\tau_{22}$. The return contribution is smaller in the high-barrier limit, because the intermediate basin is slightly deeper than the initial one. Stronger noise (larger $\kappa$) lowers these barriers, shortening the residence times and increasing the inverse MFPT [Fig.~\ref{fig:03}(d)]. The exact integral, Eq.~\eqref{eq:mfpt-exact-integral}, agrees with simulations throughout the displayed range, whereas the high-barrier approximation increasingly overestimates the inverse MFPT as $\kappa$ increases, because the effective barriers decrease from $\approx7.3$ at $\kappa=0.02$ to $\approx2.9$ at $\kappa=0.05$ and are no longer large compared with the unit diffusion scale assumed in Eq.~\eqref{eq:mfpt-integral-asymptotics}. For barriers comparable to the thermal energy in ordinary mechanical potentials, this breakdown and its remedy by the exact first-passage integral have been analyzed in the context of single-molecule pulling~\cite{Abkenar2017}.  

The physical origin of this unconventional escape dynamics lies in the stability reversal induced by negative mobility. The mechanical potential minimum becomes an effective barrier, whereas the mechanical barrier top becomes a metastable attractor. Consequently, escape is no longer governed by a single activated crossing but involves successive transitions through two effective basins, with possible returns further prolonging the process. Importantly, this additional trapping state emerges without introducing another well into the mechanical potential. Negative mobility thus changes not merely the effective barrier heights but also the stability landscape and the sequence of activated transitions governing escape.
% =====================================================
\section{Mechanical response}

Finally, we investigate how a negative-mobility particle responds to an external perturbation of the mechanical potential. Specifically, we return to the harmonic trap, $\tilde U(\tilde x)=\kappa\tilde x^2/2$, and apply an additional spatially uniform force $\tilde F$ at $\tilde t=0$, modifying the mechanical potential to $\tilde U_F(\tilde x)=\tilde U(\tilde x)-\tilde F\tilde x$. For a conventional equilibrium Brownian particle with positive mobility, this perturbation shifts the potential minimum from $\tilde x=0$ to $\tilde x=\tilde F/\kappa$, driving the particle toward the new equilibrium position along the applied force. Under negative mobility, however, the unperturbed drift $\tilde B_0(\tilde x)=\kappa\tilde x-\kappa^3\tilde x^3$ produces an effective double well, with two off-center attractors separated by a barrier. At small $\kappa$, rare switching between these attractors gives rise to long-lived positional fluctuations. This raises two questions: whether the confined ensemble responds along or against the applied force, and whether its mechanical response is governed by the same slow interwell switching that dominates spontaneous fluctuations.

\begin{figure*}[!t]
    \centering
    \includegraphics[width=\textwidth]{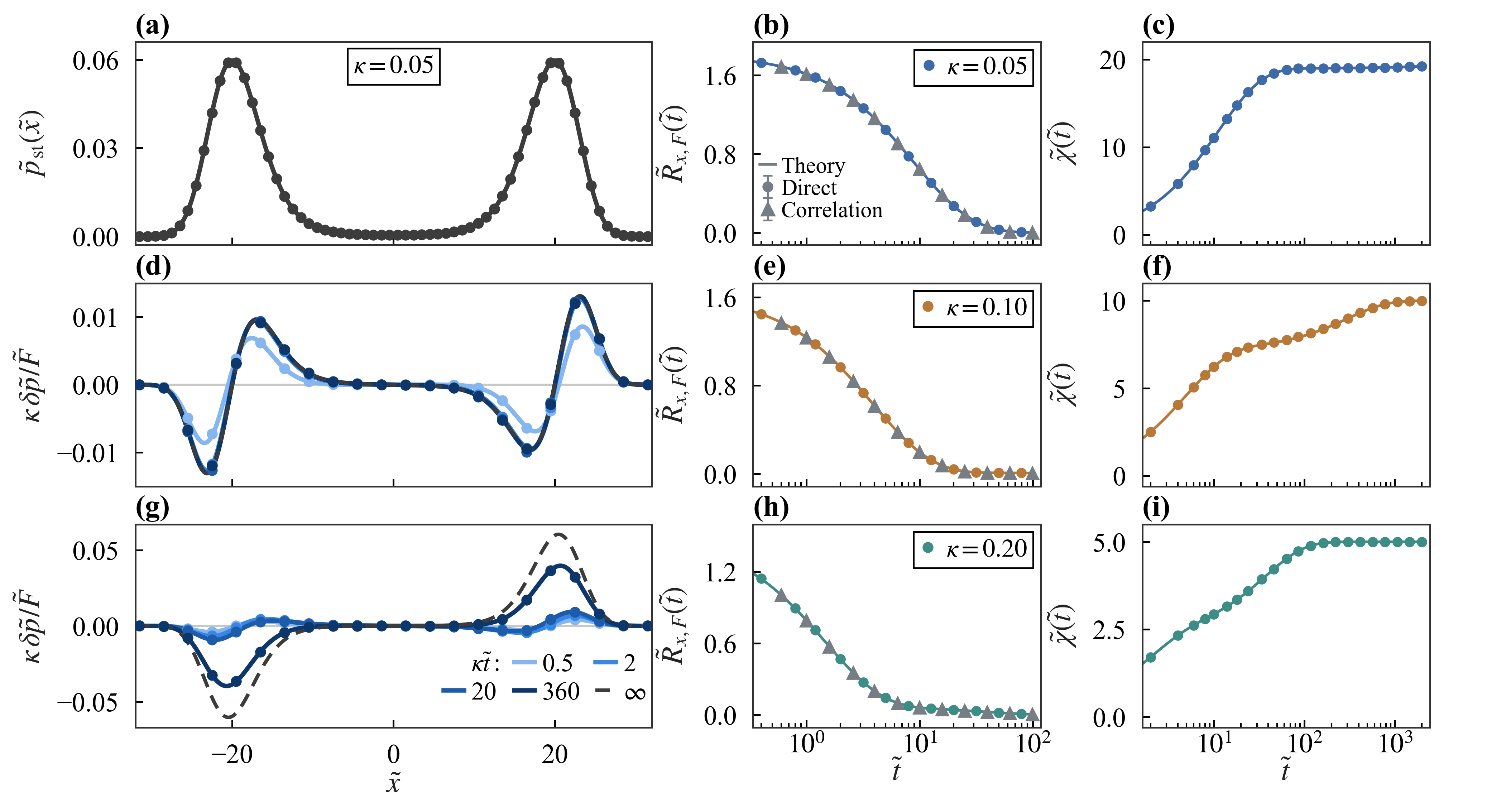}
\caption{Linear response to a uniform force $\tilde F=0.005$.
    (a) Stationary density $\tilde p_{\mathrm{st}}(\tilde x)$ at $\kappa=0.05$. (d), (g) Linear density change $\kappa\,\delta\tilde p/\tilde F$ at the same $\kappa$ for (d) the negative-mobility particle and (g) a passive unit-mobility particle in the same $\Phi$, at the times $\kappa\tilde t$ indicated in (g). Dashed lines mark the stationary limits; in (d), this limit coincides with the curves at $\kappa\tilde t=20$ and $360$.
    (b), (e), (h) Response kernel $\tilde R_{x,F}(\tilde t)$ and
    (c), (f), (i) step susceptibility $\tilde\chi(\tilde t)$ of the
    negative-mobility particle for $\kappa=0.05$, $0.10$, and $0.20$ (top to bottom).
    Lines are Fokker--Planck results [Eq.~\eqref{eq:spectral} in (b), (c), (e), (f), (h), (i)], and points are simulations. In (b), (e), and (h), circles are measured directly
    under $\pm\tilde F$, whereas triangles are reconstructed from unperturbed drift
    correlations, Eq.~\eqref{eq:response}.}
    \label{fig:04}
\end{figure*}

To quantify the response, we initialize the ensemble in the unperturbed stationary state and compare its evolution under opposite weak forces $\pm\tilde F$. We define the step susceptibility $\tilde\chi(\tilde t)=\lim_{\tilde F\to0}[\langle\tilde x(\tilde t)\rangle_{\tilde F}-\langle\tilde x(\tilde t)\rangle_{-\tilde F}]/(2\tilde F)$, where $\langle\cdot\rangle_{\tilde F}$ denotes the average under force $\tilde F$, and the corresponding response kernel $\tilde R_{x,F}(\tilde t)=\mathrm{d}\tilde\chi/\mathrm{d}\tilde t$. Numerical simulations show that $\tilde\chi(\tilde t)$ increases monotonically from zero toward a positive stationary plateau [Fig.~\ref{fig:04}(c,f,i)]. At small $\kappa$, most of the displacement develops rapidly, followed by a much slower approach to the plateau with a small remaining amplitude. Correspondingly, the response kernel remains positive throughout the observed time range and decays toward zero without changing sign [Fig.~\ref{fig:04}(b,e,h)]. Thus, the simulations reveal both a positive mechanical response and, in the high-barrier regime, a dominant displacement that develops much faster than interwell switching.

The positive stationary displacement follows directly from the geometry of harmonic confinement. Under the additional force, the drift becomes
\begin{equation}
\tilde B_F(\tilde x)=\tilde v(-\kappa\tilde x+\tilde F)
=\tilde B_0\left(\tilde x-\frac{\tilde F}{\kappa}\right).
\label{eq:bF}
\end{equation}
The effective drift potential is therefore translated, $\Phi_F(\tilde x)=\Phi(\tilde x-\tilde F/\kappa)$, up to an additive constant, and the stationary density shifts accordingly: $\tilde p_{\mathrm{st},F}(\tilde x)=\tilde p_{\mathrm{st}}(\tilde x-\tilde F/\kappa)$. Both attractors and the intervening barrier move by $\tilde F/\kappa$, while their separation, barrier height, and stationary relative populations remain unchanged. The density change is therefore a dipole within each accumulation region, $\delta\tilde p_{\mathrm{st}}\simeq-(\tilde F/\kappa)\tilde p_{\mathrm{st}}'$, with no net transfer across the translated barrier [Fig.~\ref{fig:04}(a,d)]. Although the force enters the mechanical potential as a tilt, it translates rather than tilts the effective double well. Consequently, $\langle\tilde x\rangle_{\mathrm{st},F}=\tilde F/\kappa$ and $\tilde\chi(\infty)=1/\kappa$. In dimensional units, this is the ordinary harmonic compliance $1/k$, independent of noise strength despite the negative-mobility constitutive law.

The positive static compliance does not, by itself, determine the sign of the transient response. Its initial value is controlled by the differential mobility. Expanding Eq.~\eqref{eq:bF} gives
\begin{equation}
\delta\tilde B(\tilde x)=\tilde B_F(\tilde x)-\tilde B_0(\tilde x)
=\tilde F\,\tilde v'(-\kappa\tilde x)+O(\tilde F^2),
\label{eq:db}
\end{equation}
so the perturbation probes $\tilde v'(\tilde f)=3\tilde f^2-1$ rather than the mobility $\tilde v(\tilde f)/\tilde f$. At the accumulation positions, where $|\tilde f|=1$, the differential mobility is positive, $\tilde v'(\pm1)=2$. More generally, averaging over the stationary state and integrating by parts using $\partial_{\tilde x}\tilde p_{\mathrm{st}}=\tilde B_0\tilde p_{\mathrm{st}}$ yields
\begin{equation}
\tilde R_{x,F}(0^+)
=\langle\tilde v'(-\kappa\tilde x)\rangle_{\mathrm{st}}
=\frac{\langle\tilde B_0(\tilde x)^2\rangle_{\mathrm{st}}}{\kappa}
>0.
\label{eq:R0}
\end{equation}
The initial ensemble response is therefore positive for every $\kappa>0$, even though the local drift increment near the trap center opposes the perturbation.

To determine the full transient response, we relate it to spontaneous fluctuations of the unperturbed system. The relevant fluctuation--response relation depends on the observable conjugate to the perturbation in the effective drift potential~\cite{Agarwal1972,HanggiThomas1982,Marconi2008,Prost2009}. Expanding the translated landscape gives $\delta\Phi(\tilde x)=\tilde F\tilde B_0(\tilde x)/\kappa+O(\tilde F^2)$. In this effective description, the mechanical force therefore couples to the drift rather than to position. Linearizing the Fokker--Planck equation and using the zero-current stationary condition gives
\begin{equation}
\tilde R_{x,F}(\tilde t)
=\frac{1}{\kappa}\frac{\mathrm{d}^2\tilde C}{\mathrm{d}\tilde t^2}
=\frac{1}{\kappa}
\bigl\langle\tilde B_0[\tilde x(\tilde t)]\tilde B_0[\tilde x(0)]\bigr\rangle_{\mathrm{st}},
\qquad \tilde t>0,
\label{eq:response}
\end{equation}
where $\tilde C(\tilde t)=\langle\tilde x(\tilde t)\tilde x(0)\rangle_{\mathrm{st}}$ is the stationary position autocorrelation. This relation is exact within linear response and requires neither a high-barrier nor a local Gaussian approximation. Integrating it with the stationary identity $\mathrm{d}\tilde C/\mathrm{d}\tilde t|_{\tilde t=0^+}=-1$ gives $\tilde\chi(\tilde t)=[1+\mathrm{d}\tilde C/\mathrm{d}\tilde t]/\kappa$, recovering the static susceptibility $\tilde\chi(\infty)=1/\kappa$. The kernel reconstructed from spontaneous drift correlations agrees with direct measurements [Fig.~\ref{fig:04}(b,e,h)].

Equation~\eqref{eq:response} contains contributions from both intrawell relaxation and interwell switching. To separate these processes and establish the sign of the response at all times, we expand the dynamics in eigenmodes of the backward operator $\tilde{\mathcal L}=\tilde B_0\partial_{\tilde x}+\partial_{\tilde x}^2$. The zero-current condition gives $\tilde{\mathcal L}g=\tilde p_{\mathrm{st}}^{-1}\partial_{\tilde x}(\tilde p_{\mathrm{st}}\partial_{\tilde x}g)$, which is symmetric under the stationary inner product $\langle hg\rangle_{\mathrm{st}}=\int hg\,\tilde p_{\mathrm{st}}\,\mathrm{d}\tilde x$. This symmetry expresses the reversibility of the reduced positional dynamics with respect to $\Phi$. The eigenmodes satisfy~\cite{Risken1989,Gardiner2009}
\begin{equation}
-\tilde{\mathcal L}\varphi_n=\tilde\lambda_n\varphi_n,
\qquad
\langle\varphi_n\varphi_m\rangle_{\mathrm{st}}=\delta_{nm},
\label{eq:modes}
\end{equation}
where $\tilde\lambda_0=0$ and $\varphi_0=1$ describe the stationary state, while $\tilde\lambda_n>0$ for $n\geq1$. Details of the numerical calculation are given in Sec.~S3 of the Supplemental Material~\cite{SM}. Since $\langle\tilde x\rangle_{\mathrm{st}}=0$, expanding $\tilde x$ in these modes gives
\begin{subequations}
\label{eq:spectral}
\begin{align}
\tilde C(\tilde t)
&=\sum_{n\geq1}\tilde c_n e^{-\tilde\lambda_n\tilde t},
\qquad \tilde c_n=\langle\tilde x\varphi_n\rangle_{\mathrm{st}}^2,
\label{eq:spectral-C}\\
\tilde\chi(\tilde t)
&=\frac{1}{\kappa}
\sum_{n\geq1}\rho_n(1-e^{-\tilde\lambda_n\tilde t}),
\qquad \rho_n=\tilde\lambda_n\tilde c_n,
\label{eq:spectral-chi}\\
\tilde R_{x,F}(\tilde t)
&=\frac{1}{\kappa}
\sum_{n\geq1}\tilde\lambda_n\rho_n e^{-\tilde\lambda_n\tilde t}.
\label{eq:spectral-R}
\end{align}
\end{subequations}
Here, $\tilde c_n$ measures the contribution of mode $n$ to the positional variance, with $\sum_n\tilde c_n=\langle\tilde x^2\rangle_{\mathrm{st}}$, whereas $\rho_n$ represents its fraction of the final displacement, with $\sum_n\rho_n=1$ following from $\mathrm{d}\tilde C/\mathrm{d}\tilde t|_{\tilde t=0^+}=-1$.

The spectral representation resolves both questions posed above. First, all weights in Eq.~\eqref{eq:spectral-R} are nonnegative, establishing $\tilde R_{x,F}(\tilde t)>0$ at every finite time and a monotonic increase of $\tilde\chi(\tilde t)$ toward $1/\kappa$. The positive response therefore follows from reversibility together with the translational coupling of the perturbation, rather than from a positive local mobility. Second, spontaneous fluctuations and mechanical response share the same relaxation rates but assign different weights to the corresponding modes. The additional factor $\tilde\lambda_n$ in $\rho_n=\tilde\lambda_n\tilde c_n$ suppresses the contribution of slowly relaxing modes to the displacement, allowing the response to develop predominantly through faster relaxation processes.

At $\kappa=0.05$, for example, the slow mode ($n=1$), which describes interwell switching, accounts for approximately $97\%$ of the positional variance but only $5\%$ of the final displacement [Table~S1]. Most of the displacement is instead carried by the odd modes $n=3$ and $n=5$, with $\rho_3\simeq0.58$ and $\rho_5\simeq0.30$, whose relaxation rates are comparable to the intrawell rate $2\kappa$. The displacement therefore develops mainly through intrawell relaxation, while the slow mode only sets the long-time tail, which keeps $\tilde\chi$ slightly below $1/\kappa$ within the window of Fig.~\ref{fig:04}(c). As $\kappa$ increases, the lower effective barrier accelerates interwell switching and raises $\rho_1$ to approximately $0.54$ at $\kappa=0.20$ [Table~S1], so intrawell dominance is characteristic of the high-barrier regime. Because the slow mode nonetheless dominates positional fluctuations, its suppression in the displacement must originate either from the effective double well itself or from how the force couples to it.

The distinction between fluctuations and response becomes particularly clear by comparison with a passive unit-mobility particle moving in the same effective drift potential $\Phi$, with the same diffusion coefficient and hence identical unperturbed positional dynamics. An additional uniform force $\tilde F$ then enters as an additive drift, $\tilde B_0\to\tilde B_0+\tilde F$, and tilts the potential, $\Phi\to\Phi-\tilde F\tilde x$. Its response is $\tilde R^{\mathrm p}(\tilde t)=-\mathrm{d}\tilde C/\mathrm{d}\tilde t$ and $\tilde\chi^{\mathrm p}(\tilde t)=\tilde C(0)-\tilde C(\tilde t)$, so the normalized step response has the same modal weights as the normalized positional correlation [Table~S2]. At $\kappa=0.05$, $\tilde\chi^{\mathrm p}(\infty)=\tilde C(0)=\langle\tilde x^2\rangle_{\mathrm{st}}\simeq374.7$, nearly 19 times $\tilde\chi(\infty)=1/\kappa=20$ for the negative-mobility particle. The passive particle's displacement therefore requires probability transfer from the left to the right well~\cite{Gammaitoni1998}, which remains incomplete even at $\kappa\tilde t=360\simeq\kappa/\tilde\lambda_1$ [Fig.~\ref{fig:04}(g)]. For the negative-mobility particle, by contrast, the density change remains a dipole within each well and is already close to its steady state by $\kappa\tilde t\approx 20$ [Fig.~\ref{fig:04}(d)]. This difference reflects how the force perturbs the dynamics, rather than broken detailed balance or a failure of fluctuation--response theory~\cite{Baiesi2013,Dembo2010}. Thus, under harmonic confinement, negative mobility can generate long-lived positional fluctuations while allowing most of the positive mechanical displacement to develop on the much shorter intrawell timescale.

% =====================================================
\section{Discussion}

Negative mobility changes the correspondence between mechanical potential landscapes and stochastic stability. In a harmonic trap, weak restoring forces drive particles away from the mechanical minimum, while stronger forces restore confinement, producing two accumulation regions at the force-reversal positions. The same mechanism converts the barrier top of a cubic well into an intermediate metastable state. Reaching the mechanical barrier therefore no longer marks departure from a trapped state; it leads instead to further trapping. In the high-barrier regime, the escape time contains two activated residence contributions, supplemented by returns to the initial basin. Negative mobility thus changes the sequence of metastable states encountered during escape, rather than merely modifying a conventional barrier height. Both effects arise without adding wells to the mechanical potential, showing that its geometry must be considered together with the force--velocity relation to predict localization and escape.

The mechanical response reveals a complementary contrast: stability reversal does not imply response reversal. Under harmonic confinement, a uniform force translates rather than tilts the effective double well, preserving its shape and stationary relative basin populations. The ensemble consequently retains the ordinary dimensional harmonic compliance, while reversibility and the translational coupling ensure a positive response kernel at all times. At small $\kappa$, most of the displacement develops through intrawell relaxation, even though interwell switching dominates positional fluctuations. This distinction does not involve separate relaxation spectra: fluctuations and response share the same modes, but the response weights contain an additional factor of the relaxation rate. The slow switching mode therefore contributes little to the displacement while still controlling its ultimate long-time tail. A long positional correlation time need not indicate the timescale on which most of the mechanical response develops.

These findings also clarify what can be inferred from unperturbed motion. The confined positional process remains reversible with respect to $\Phi$ and can be reproduced by a passive diffusion in the same effective drift potential. Stationary densities and spontaneous positional correlations alone therefore cannot distinguish these two descriptions~\cite{Fodor2016}. Their responses differ because a uniform force acts differently: it tilts the potential of the matched passive particle but translates the effective drift potential of the negative-mobility particle. Tilting transfers probability between the wells, whereas translation shifts each population within its own well. The distinction lies in the physical force coupling, not in broken detailed balance or a failure of fluctuation--response theory. In particular, a positive measured susceptibility does not exclude negative mobility, just as effective bistability does not imply a switching-dominated displacement. Joint measurements of localization, spontaneous fluctuations, and force response would provide a more informative characterization than any one observable alone~\cite{Martin2001,Mizuno2007}.

The derivation also identifies which conclusions extend beyond the chosen cubic law. A smooth local force--velocity relation with $v(0)=0$ and $v'(0)<0$ reverses the stability of nondegenerate mechanical extrema, while harmonic confinement gives $\tilde B_F(\tilde x)=\tilde B_0(\tilde x-\tilde F/\kappa)$ for any instantaneous local constitutive law. Quantitative barrier heights, stationary moments, and response weights nevertheless depend on the specific model, and the positivity argument additionally requires the reversible stationary dynamics considered here. Memory, force-dependent diffusion, and anharmonic confinement require separate analysis. Moreover, the effective equilibrium representation of the reduced position dynamics does not resolve the nonequilibrium mechanism or energetic cost of sustaining negative mobility. Within these limits, the central implication is that mechanical landscapes organize stochastic motion only through the constitutive response they induce: the effective drift determines stability and escape, while its coupling to an additional force determines how those states respond.

% ========================================================
\begin{acknowledgments}
This work is supported by the National Key Research and Development Program of China (No. 2023YFA1407500), the National Natural Science Foundation of China No. 12374219, the 111 project (B16029), and Fujian Provincial Excellent Postdoctoral Program.
\end{acknowledgments}

\bibliographystyle{apsrev4-1}
\bibliography{negative_mobility}

@article{Einstein1905,
  author  = {Einstein, A.},
  title   = {{\"U}ber die von der molekularkinetischen {T}heorie der {W}{\"a}rme geforderte {B}ewegung von in ruhenden {F}l{\"u}ssigkeiten suspendierten {T}eilchen},
  journal = {Ann. Phys. (Leipzig)},
  volume  = {322},
  number  = {8},
  pages   = {549--560},
  year    = {1905},
  doi     = {10.1002/andp.19053220806}
}

@article{Kramers1940,
  author  = {Kramers, H. A.},
  title   = {{B}rownian motion in a field of force and the diffusion model of chemical reactions},
  journal = {Physica},
  volume  = {7},
  number  = {4},
  pages   = {284--304},
  year    = {1940},
  doi     = {10.1016/S0031-8914(40)90098-2}
}

@article{Hanggi1990,
  author  = {H{\"a}nggi, P. and Talkner, P. and Borkovec, M.},
  title   = {Reaction-rate theory: fifty years after {K}ramers},
  journal = {Rev. Mod. Phys.},
  volume  = {62},
  number  = {2},
  pages   = {251--341},
  year    = {1990},
  doi     = {10.1103/RevModPhys.62.251}
}

@article{Kubo1966,
  author  = {Kubo, R.},
  title   = {The fluctuation-dissipation theorem},
  journal = {Rep. Prog. Phys.},
  volume  = {29},
  number  = {1},
  pages   = {255--284},
  year    = {1966},
  doi     = {10.1088/0034-4885/29/1/306}
}

@article{Marconi2008,
  author  = {{Marini Bettolo Marconi}, U. and Puglisi, A. and Rondoni, L. and Vulpiani, A.},
  title   = {Fluctuation--dissipation: {R}esponse theory in statistical physics},
  journal = {Phys. Rep.},
  volume  = {461},
  number  = {4--6},
  pages   = {111--195},
  year    = {2008},
  doi     = {10.1016/j.physrep.2008.02.002}
}

@article{Marchetti2013,
  author  = {Marchetti, M. C. and Joanny, J. F. and Ramaswamy, S. and Liverpool, T. B. and Prost, J. and Rao, M. and Simha, R. A.},
  title   = {Hydrodynamics of soft active matter},
  journal = {Rev. Mod. Phys.},
  volume  = {85},
  number  = {3},
  pages   = {1143--1189},
  year    = {2013},
  doi     = {10.1103/RevModPhys.85.1143}
}

@article{Bechinger2016,
  author  = {Bechinger, C. and {Di Leonardo}, R. and L{\"o}wen, H. and Reichhardt, C. and Volpe, G. and Volpe, G.},
  title   = {Active particles in complex and crowded environments},
  journal = {Rev. Mod. Phys.},
  volume  = {88},
  number  = {4},
  pages   = {045006},
  year    = {2016},
  doi     = {10.1103/RevModPhys.88.045006}
}

@article{Harada2005,
  author  = {Harada, T. and Sasa, S.-i.},
  title   = {Equality connecting energy dissipation with a violation of the fluctuation-response relation},
  journal = {Phys. Rev. Lett.},
  volume  = {95},
  number  = {13},
  pages   = {130602},
  year    = {2005},
  doi     = {10.1103/PhysRevLett.95.130602}
}

@article{Wu2000,
  author  = {Wu, X.-L. and Libchaber, A.},
  title   = {Particle diffusion in a quasi-two-dimensional bacterial bath},
  journal = {Phys. Rev. Lett.},
  volume  = {84},
  number  = {13},
  pages   = {3017--3020},
  year    = {2000},
  doi     = {10.1103/PhysRevLett.84.3017}
}

@article{Maggi2014,
  author  = {Maggi, C. and Paoluzzi, M. and Pellicciotta, N. and Lepore, A. and Angelani, L. and {Di Leonardo}, R.},
  title   = {Generalized energy equipartition in harmonic oscillators driven by active baths},
  journal = {Phys. Rev. Lett.},
  volume  = {113},
  number  = {23},
  pages   = {238303},
  year    = {2014},
  doi     = {10.1103/PhysRevLett.113.238303}
}

@article{Krueger2009,
  author  = {Kr{\"u}ger, M. and Fuchs, M.},
  title   = {Fluctuation dissipation relations in stationary states of interacting {B}rownian particles under shear},
  journal = {Phys. Rev. Lett.},
  volume  = {102},
  number  = {13},
  pages   = {135701},
  year    = {2009},
  doi     = {10.1103/PhysRevLett.102.135701}
}

@article{Baiesi2009,
  author  = {Baiesi, M. and Maes, C. and Wynants, B.},
  title   = {Fluctuations and response of nonequilibrium states},
  journal = {Phys. Rev. Lett.},
  volume  = {103},
  number  = {1},
  pages   = {010602},
  year    = {2009},
  doi     = {10.1103/PhysRevLett.103.010602}
}

@article{Seifert2010,
  author  = {Seifert, U. and Speck, T.},
  title   = {Fluctuation-dissipation theorem in nonequilibrium steady states},
  journal = {Europhys. Lett.},
  volume  = {89},
  number  = {1},
  pages   = {10007},
  year    = {2010},
  doi     = {10.1209/0295-5075/89/10007}
}

@article{Baiesi2013,
  author  = {Baiesi, M. and Maes, C.},
  title   = {An update on the nonequilibrium linear response},
  journal = {New J. Phys.},
  volume  = {15},
  number  = {1},
  pages   = {013004},
  year    = {2013},
  doi     = {10.1088/1367-2630/15/1/013004}
}

@article{Dembo2010,
  author  = {Dembo, A. and Deuschel, J.-D.},
  title   = {{M}arkovian perturbation, response and fluctuation dissipation theorem},
  journal = {Ann. Inst. H. Poincar{\'e} Probab. Statist.},
  volume  = {46},
  number  = {3},
  pages   = {822--852},
  year    = {2010},
  doi     = {10.1214/10-AIHP370},
  eprint  = {0710.4394},
  archivePrefix = {arXiv}
}

@article{Agarwal1972,
  author  = {Agarwal, G. S.},
  title   = {Fluctuation-dissipation theorems for systems in non-thermal equilibrium and applications},
  journal = {Z. Phys.},
  volume  = {252},
  number  = {1},
  pages   = {25--38},
  year    = {1972},
  doi     = {10.1007/BF01391621}
}

@article{HanggiThomas1982,
  author  = {H{\"a}nggi, P. and Thomas, H.},
  title   = {Stochastic processes: {T}ime evolution, symmetries and linear response},
  journal = {Phys. Rep.},
  volume  = {88},
  number  = {4},
  pages   = {207--319},
  year    = {1982},
  doi     = {10.1016/0370-1573(82)90045-X}
}

@article{Blickle2007,
  author  = {Blickle, V. and Speck, T. and Lutz, C. and Seifert, U. and Bechinger, C.},
  title   = {{E}instein relation generalized to nonequilibrium},
  journal = {Phys. Rev. Lett.},
  volume  = {98},
  number  = {21},
  pages   = {210601},
  year    = {2007},
  doi     = {10.1103/PhysRevLett.98.210601}
}

@article{Prost2009,
  author  = {Prost, J. and Joanny, J.-F. and Parrondo, J. M. R.},
  title   = {Generalized fluctuation-dissipation theorem for steady-state systems},
  journal = {Phys. Rev. Lett.},
  volume  = {103},
  number  = {9},
  pages   = {090601},
  year    = {2009},
  doi     = {10.1103/PhysRevLett.103.090601}
}

@article{Argun2016,
  author  = {Argun, A. and Moradi, A.-R. and Pin{\c{c}}e, E. and Bagci, G. B. and Imparato, A. and Volpe, G.},
  title   = {Non-{B}oltzmann stationary distributions and nonequilibrium relations in active baths},
  journal = {Phys. Rev. E},
  volume  = {94},
  number  = {6},
  pages   = {062150},
  year    = {2016},
  doi     = {10.1103/PhysRevE.94.062150}
}

@article{Fodor2016,
  author  = {Fodor, {\'E}. and Nardini, C. and Cates, M. E. and Tailleur, J. and Visco, P. and {van Wijland}, F.},
  title   = {How far from equilibrium is active matter?},
  journal = {Phys. Rev. Lett.},
  volume  = {117},
  number  = {3},
  pages   = {038103},
  year    = {2016},
  doi     = {10.1103/PhysRevLett.117.038103}
}

@article{Martin2001,
  author  = {Martin, P. and Hudspeth, A. J. and J{\"u}licher, F.},
  title   = {Comparison of a hair bundle's spontaneous oscillations with its response to mechanical stimulation reveals the underlying active process},
  journal = {Proc. Natl. Acad. Sci. USA},
  volume  = {98},
  number  = {25},
  pages   = {14380--14385},
  year    = {2001},
  doi     = {10.1073/pnas.251530598}
}

@article{Mizuno2007,
  author  = {Mizuno, D. and Tardin, C. and Schmidt, C. F. and MacKintosh, F. C.},
  title   = {Nonequilibrium mechanics of active cytoskeletal networks},
  journal = {Science},
  volume  = {315},
  number  = {5810},
  pages   = {370--373},
  year    = {2007},
  doi     = {10.1126/science.1134404}
}

@article{Eichhorn2002,
  author  = {Eichhorn, R. and Reimann, P. and H{\"a}nggi, P.},
  title   = {{B}rownian motion exhibiting absolute negative mobility},
  journal = {Phys. Rev. Lett.},
  volume  = {88},
  number  = {19},
  pages   = {190601},
  year    = {2002},
  doi     = {10.1103/PhysRevLett.88.190601}
}

@article{Machura2007,
  author  = {Machura, L. and Kostur, M. and Talkner, P. and {\L}uczka, J. and H{\"a}nggi, P.},
  title   = {Absolute negative mobility induced by thermal equilibrium fluctuations},
  journal = {Phys. Rev. Lett.},
  volume  = {98},
  number  = {4},
  pages   = {040601},
  year    = {2007},
  doi     = {10.1103/PhysRevLett.98.040601}
}

@article{Sarracino2016,
  author  = {Sarracino, A. and Cecconi, F. and Puglisi, A. and Vulpiani, A.},
  title   = {Nonlinear response of inertial tracers in steady laminar flows: {D}ifferential and absolute negative mobility},
  journal = {Phys. Rev. Lett.},
  volume  = {117},
  number  = {17},
  pages   = {174501},
  year    = {2016},
  doi     = {10.1103/PhysRevLett.117.174501}
}

@article{Ros2005,
  author  = {Ros, A. and Eichhorn, R. and Regtmeier, J. and Duong, T. T. and Reimann, P. and Anselmetti, D.},
  title   = {Absolute negative particle mobility},
  journal = {Nature (London)},
  volume  = {436},
  pages   = {928},
  year    = {2005},
  doi     = {10.1038/436928a}
}

@article{Nagel2008,
  author  = {Nagel, J. and Speer, D. and Gaber, T. and Sterck, A. and Eichhorn, R. and Reimann, P. and Ilin, K. and Siegel, M. and Koelle, D. and Kleiner, R.},
  title   = {Observation of negative absolute resistance in a {J}osephson junction},
  journal = {Phys. Rev. Lett.},
  volume  = {100},
  number  = {21},
  pages   = {217001},
  year    = {2008},
  doi     = {10.1103/PhysRevLett.100.217001}
}

@article{Foffano2012,
  author  = {Foffano, G. and Lintuvuori, J. S. and Stratford, K. and Cates, M. E. and Marenduzzo, D.},
  title   = {Colloids in active fluids: {A}nomalous microrheology and negative drag},
  journal = {Phys. Rev. Lett.},
  volume  = {109},
  number  = {2},
  pages   = {028103},
  year    = {2012},
  doi     = {10.1103/PhysRevLett.109.028103}
}

@article{Cecconi2018,
  author  = {Cecconi, F. and Puglisi, A. and Sarracino, A. and Vulpiani, A.},
  title   = {Anomalous mobility of a driven active particle in a steady laminar flow},
  journal = {J. Phys.: Condens. Matter},
  volume  = {30},
  number  = {26},
  pages   = {264002},
  year    = {2018},
  doi     = {10.1088/1361-648X/aac4f0}
}

@article{Rizkallah2023,
  author  = {Rizkallah, P. and Sarracino, A. and B{\'e}nichou, O. and Illien, P.},
  title   = {Absolute negative mobility of an active tracer in a crowded environment},
  journal = {Phys. Rev. Lett.},
  volume  = {130},
  number  = {21},
  pages   = {218201},
  year    = {2023},
  doi     = {10.1103/PhysRevLett.130.218201}
}

@article{Cleuren2001,
  author  = {Cleuren, B. and {Van den Broeck}, C.},
  title   = {Ising model for a {B}rownian donkey},
  journal = {Europhys. Lett.},
  volume  = {54},
  number  = {1},
  pages   = {1--6},
  year    = {2001},
  doi     = {10.1209/epl/i2001-00274-6}
}

@article{Ghosh2014,
  author  = {Ghosh, P. K. and H{\"a}nggi, P. and Marchesoni, F. and Nori, F.},
  title   = {Giant negative mobility of {J}anus particles in a corrugated channel},
  journal = {Phys. Rev. E},
  volume  = {89},
  number  = {6},
  pages   = {062115},
  year    = {2014},
  doi     = {10.1103/PhysRevE.89.062115}
}

@article{Hanggi2009,
  author  = {H{\"a}nggi, P. and Marchesoni, F.},
  title   = {Artificial {B}rownian motors: {C}ontrolling transport on the nanoscale},
  journal = {Rev. Mod. Phys.},
  volume  = {81},
  number  = {1},
  pages   = {387--442},
  year    = {2009},
  doi     = {10.1103/RevModPhys.81.387}
}

@article{Benichou2014,
  author  = {B{\'e}nichou, O. and Illien, P. and Oshanin, G. and Sarracino, A. and Voituriez, R.},
  title   = {Microscopic theory for negative differential mobility in crowded environments},
  journal = {Phys. Rev. Lett.},
  volume  = {113},
  number  = {26},
  pages   = {268002},
  year    = {2014},
  doi     = {10.1103/PhysRevLett.113.268002}
}

@article{Reichhardt2018,
  author  = {Reichhardt, C. and Reichhardt, C. J. O.},
  title   = {Negative differential mobility and trapping in active matter systems},
  journal = {J. Phys.: Condens. Matter},
  volume  = {30},
  number  = {1},
  pages   = {015404},
  year    = {2018},
  doi     = {10.1088/1361-648X/aa9c5f}
}

@article{Eichhorn2002PRE,
  author  = {Eichhorn, R. and Reimann, P. and H{\"a}nggi, P.},
  title   = {Paradoxical motion of a single {B}rownian particle: {A}bsolute negative mobility},
  journal = {Phys. Rev. E},
  volume  = {66},
  number  = {6},
  pages   = {066132},
  year    = {2002},
  doi     = {10.1103/PhysRevE.66.066132}
}

@article{Slapik2019,
  author  = {S{\l}apik, A. and {\L}uczka, J. and H{\"a}nggi, P. and Spiechowicz, J.},
  title   = {Tunable mass separation via negative mobility},
  journal = {Phys. Rev. Lett.},
  volume  = {122},
  number  = {7},
  pages   = {070602},
  year    = {2019},
  doi     = {10.1103/PhysRevLett.122.070602}
}

@article{Zia2002,
  author  = {Zia, R. K. P. and Praestgaard, E. L. and Mouritsen, O. G.},
  title   = {Getting more from pushing less: {N}egative specific heat and conductivity in nonequilibrium steady states},
  journal = {Am. J. Phys.},
  volume  = {70},
  number  = {4},
  pages   = {384--392},
  year    = {2002},
  doi     = {10.1119/1.1427088}
}

@article{Leitmann2013,
  author  = {Leitmann, S. and Franosch, T.},
  title   = {Nonlinear response in the driven lattice {L}orentz gas},
  journal = {Phys. Rev. Lett.},
  volume  = {111},
  number  = {19},
  pages   = {190603},
  year    = {2013},
  doi     = {10.1103/PhysRevLett.111.190603}
}

@article{Tailleur2009,
  author  = {Tailleur, J. and Cates, M. E.},
  title   = {Sedimentation, trapping, and rectification of dilute bacteria},
  journal = {Europhys. Lett.},
  volume  = {86},
  number  = {6},
  pages   = {60002},
  year    = {2009},
  doi     = {10.1209/0295-5075/86/60002}
}

@article{Takatori2016,
  author  = {Takatori, S. C. and {De Dier}, R. and Vermant, J. and Brady, J. F.},
  title   = {Acoustic trapping of active matter},
  journal = {Nat. Commun.},
  volume  = {7},
  pages   = {10694},
  year    = {2016},
  doi     = {10.1038/ncomms10694}
}

@article{Malakar2020,
  author  = {Malakar, K. and Das, A. and Kundu, A. and Kumar, K. V. and Dhar, A.},
  title   = {Steady state of an active {B}rownian particle in a two-dimensional harmonic trap},
  journal = {Phys. Rev. E},
  volume  = {101},
  number  = {2},
  pages   = {022610},
  year    = {2020},
  doi     = {10.1103/PhysRevE.101.022610}
}

@article{Pototsky2012,
  author  = {Pototsky, A. and Stark, H.},
  title   = {Active {B}rownian particles in two-dimensional traps},
  journal = {Europhys. Lett.},
  volume  = {98},
  number  = {5},
  pages   = {50004},
  year    = {2012},
  doi     = {10.1209/0295-5075/98/50004}
}

@article{Dhar2019,
  author  = {Dhar, A. and Kundu, A. and Majumdar, S. N. and Sabhapandit, S. and Schehr, G.},
  title   = {Run-and-tumble particle in one-dimensional confining potentials: {S}teady-state, relaxation, and first-passage properties},
  journal = {Phys. Rev. E},
  volume  = {99},
  number  = {3},
  pages   = {032132},
  year    = {2019},
  doi     = {10.1103/PhysRevE.99.032132}
}

@article{Caprini2019,
  author  = {Caprini, L. and {Marini Bettolo Marconi}, U. and Puglisi, A. and Vulpiani, A.},
  title   = {Active escape dynamics: {T}he effect of persistence on barrier crossing},
  journal = {J. Chem. Phys.},
  volume  = {150},
  number  = {2},
  pages   = {024902},
  year    = {2019},
  doi     = {10.1063/1.5080537}
}

@article{Militaru2021,
  author  = {Militaru, A. and Innerbichler, M. and Frimmer, M. and Tebbenjohanns, F. and Novotny, L. and Dellago, C.},
  title   = {Escape dynamics of active particles in multistable potentials},
  journal = {Nat. Commun.},
  volume  = {12},
  pages   = {2446},
  year    = {2021},
  doi     = {10.1038/s41467-021-22647-6}
}

@article{Geiseler2016,
  author  = {Geiseler, A. and H{\"a}nggi, P. and Schmid, G.},
  title   = {{K}ramers escape of a self-propelled particle},
  journal = {Eur. Phys. J. B},
  volume  = {89},
  number  = {8},
  pages   = {175},
  year    = {2016},
  doi     = {10.1140/epjb/e2016-70359-0}
}

@article{Woillez2019,
  author  = {Woillez, E. and Zhao, Y. and Kafri, Y. and Lecomte, V. and Tailleur, J.},
  title   = {Activated escape of a self-propelled particle from a metastable state},
  journal = {Phys. Rev. Lett.},
  volume  = {122},
  number  = {25},
  pages   = {258001},
  year    = {2019},
  doi     = {10.1103/PhysRevLett.122.258001}
}

@article{Bovier2004,
  author  = {Bovier, A. and Eckhoff, M. and Gayrard, V. and Klein, M.},
  title   = {Metastability in reversible diffusion processes {I}: {S}harp asymptotics for capacities and exit times},
  journal = {J. Eur. Math. Soc.},
  volume  = {6},
  number  = {4},
  pages   = {399--424},
  year    = {2004},
  doi     = {10.4171/JEMS/14}
}

@article{Gammaitoni1998,
  author  = {Gammaitoni, L. and H{\"a}nggi, P. and Jung, P. and Marchesoni, F.},
  title   = {Stochastic resonance},
  journal = {Rev. Mod. Phys.},
  volume  = {70},
  number  = {1},
  pages   = {223--287},
  year    = {1998},
  doi     = {10.1103/RevModPhys.70.223}
}

@book{Risken1989,
  author    = {Risken, H.},
  title     = {The {F}okker--{P}lanck Equation: {M}ethods of Solution and Applications},
  edition   = {2nd},
  series    = {Springer Series in Synergetics},
  volume    = {18},
  publisher = {Springer},
  address   = {Berlin},
  year      = {1996},
  doi       = {10.1007/978-3-642-61544-3}
}

@book{Gardiner2009,
  author    = {Gardiner, C.},
  title     = {Stochastic Methods: {A} Handbook for the Natural and Social Sciences},
  edition   = {4th},
  series    = {Springer Series in Synergetics},
  volume    = {13},
  publisher = {Springer},
  address   = {Berlin},
  year      = {2009}
}

@book{Redner2001,
  author    = {Redner, S.},
  title     = {A Guide to First-Passage Processes},
  publisher = {Cambridge University Press},
  address   = {Cambridge},
  year      = {2001},
  doi       = {10.1017/CBO9780511606014}
}

@book{Kloeden1992,
  author    = {Kloeden, P. E. and Platen, E.},
  title     = {Numerical Solution of Stochastic Differential Equations},
  publisher = {Springer},
  address   = {Berlin},
  year      = {1992},
  doi       = {10.1007/978-3-662-12616-5}
}

@article{Abkenar2017,
  author  = {Abkenar, M. and Gray, T. H. and Zaccone, A.},
  title   = {Dissociation rates from single-molecule pulling experiments under large thermal fluctuations or large applied force},
  journal = {Phys. Rev. E},
  volume  = {95},
  number  = {4},
  pages   = {042413},
  year    = {2017},
  doi     = {10.1103/PhysRevE.95.042413}
}

@misc{SM,
  note = {See Supplemental Material at [URL will be inserted by publisher] for the weak- and strong-noise expansions of the stationary moments and the numerical spectral expansion.}
}
\end{document}

\title{Supplemental Material for ``Stability reversal, activated escape, and mechanical response of negative-mobility particles''}

\author{Junteng Li}
%\thanks{Corresponding author: zhyou@xmu.edu.cn}
\affiliation{Fujian Provincial Key Laboratory for Soft Functional Materials Research,
Research Institute for Biomimetics and Soft Matter, Department of
Physics, Xiamen University, Xiamen, Fujian 361005, China}

\author{Zhihong You}
\thanks{Corresponding author: zhyou@xmu.edu.cn}
\affiliation{Fujian Provincial Key Laboratory for Soft Functional Materials Research,
Research Institute for Biomimetics and Soft Matter, Department of
Physics, Xiamen University, Xiamen, Fujian 361005, China}
\date{\today}

\maketitle

\section{Stationary moments in the weak- and strong-noise limits}
\label{app:width}

Throughout this Supplemental Material, we use the dimensionless units of the main text: lengths are measured in $\ell_0=D/v_0$, times in $\tau_0=D/v_0^2$, and forces in $f_0$, so that the diffusion coefficient equals unity. As in the main text, tildes mark quantities expressed in these units, such as $\tilde x$, $\tilde t$, $\tilde F$, $\tilde p_{\mathrm{st}}$, $\tilde B_0$, $\tilde\sigma_+^2$, $\tilde\lambda_n$, and $\tilde c_n$, whereas intrinsically dimensionless quantities, such as $\Phi$, $\kappa$, and $\rho_n$, carry no tilde. The dimensionless trap stiffness is $\kappa=kD/(v_0f_0)$, with $k$ the dimensional stiffness. At fixed $k$, $v_0$, and $f_0$, varying $\kappa$ therefore corresponds to varying the noise strength $D$.

For the harmonic trap, the effective drift potential
\begin{equation}
\Phi(\tilde x)=\frac{(\kappa^2\tilde x^2-1)^2}{4\kappa}
\label{eq:app_Phi}
\end{equation}
has minima at the attractors $\tilde x_\pm=\pm\kappa^{-1}$, separated by the barrier $\Delta\Phi=1/(4\kappa)$. Because $\Phi$ is even, the two accumulation regions are mirror images, and it suffices to consider $\tilde x>0$. The width of the right population is the conditional variance
\begin{equation}
\tilde\sigma_+^2=\langle\tilde x^2\rangle_+-\langle\tilde x\rangle_+^2,
\qquad
\langle g\rangle_+=\frac{\int_0^\infty g(\tilde x)\,e^{-\Phi(\tilde x)}\,\mathrm d\tilde x}{\int_0^\infty e^{-\Phi(\tilde x)}\,\mathrm d\tilde x},
\label{eq:app_sigma}
\end{equation}
and $\langle\tilde x^2\rangle_{\rm st}=\langle\tilde x^2\rangle_+$ by symmetry. This expression is exact for all $\kappa$. Its limiting forms follow from approximating $e^{-\Phi}$ over the region sampled by the particles.

\textit{Weak noise, $\kappa\ll1$.}---In this limit, $\Delta\Phi\gg1$ and the particles remain close to the attractor. With $\xi=\tilde x-\kappa^{-1}$, effective drift potential expands exactly as
\begin{equation}
\Phi=\kappa\,\xi^2+\kappa^2\xi^3+\tfrac14\kappa^3\xi^4 .
\label{eq:app_expand}
\end{equation}
The quadratic term confines the particles to $|\xi|\sim\kappa^{-1/2}$, where the cubic and quartic terms are of order $\kappa^{1/2}$ and $\kappa$, respectively. To leading order, the stationary weight is therefore
\begin{equation}
e^{-\Phi}\simeq e^{-\kappa\xi^2}\left(1-\kappa^2\xi^3\right).
\label{eq:app_gauss}
\end{equation}
Substituting Eq.~\eqref{eq:app_gauss} into Eq.~\eqref{eq:app_sigma} gives
\begin{subequations}
\label{eq:app_xi}
\begin{align}
\langle\xi\rangle_+&\simeq
\frac{\int\xi\,e^{-\kappa\xi^2}\,\mathrm d\xi-\kappa^2\int\xi^4e^{-\kappa\xi^2}\,\mathrm d\xi}
{\int e^{-\kappa\xi^2}\,\mathrm d\xi-\kappa^2\int\xi^3e^{-\kappa\xi^2}\,\mathrm d\xi}
=-\kappa^2\cdot\frac{3}{4\kappa^2}=-\frac34,
\label{eq:app_xi_mean}\\[4pt]
\langle\xi^2\rangle_+&\simeq
\frac{\int\xi^2e^{-\kappa\xi^2}\,\mathrm d\xi-\kappa^2\int\xi^5e^{-\kappa\xi^2}\,\mathrm d\xi}
{\int e^{-\kappa\xi^2}\,\mathrm d\xi-\kappa^2\int\xi^3e^{-\kappa\xi^2}\,\mathrm d\xi}
=\frac{1}{2\kappa},
\label{eq:app_xi_var}
\end{align}
\end{subequations}
where the locally expanded integrals, defined for $\xi>-\kappa^{-1}$, are extended to the full real line with corrections of order $e^{-1/\kappa}$ only. The integrals of odd powers of $\xi$ then vanish, and the remaining Gaussian ratios are $\int\xi^2e^{-\kappa\xi^2}\mathrm d\xi/\int e^{-\kappa\xi^2}\mathrm d\xi=1/(2\kappa)$ and $\int\xi^4e^{-\kappa\xi^2}\mathrm d\xi/\int e^{-\kappa\xi^2}\mathrm d\xi=3/(4\kappa^2)$. The cubic term thus shifts the mean inward but, at this order, leaves the width unchanged. Using $\langle\tilde x\rangle_+=\kappa^{-1}+\langle\xi\rangle_+$, $\tilde\sigma_+^2=\langle\xi^2\rangle_+-\langle\xi\rangle_+^2$, and $\langle\tilde x^2\rangle_{\rm st}=\kappa^{-2}+2\kappa^{-1}\langle\xi\rangle_++\langle\xi^2\rangle_+$, and measuring lengths relative to the attractor distance $\tilde x_+=\kappa^{-1}$, we obtain
\begin{equation}
\label{eq:weak-moments}
\begin{aligned}
\kappa\langle\tilde x\rangle_+&=1-\tfrac34\kappa+O(\kappa^2),\\
\kappa^2\tilde\sigma_+^2&=\tfrac12\kappa+O(\kappa^2),\\
\kappa^2\langle\tilde x^2\rangle_{\rm st}&=1-\kappa+O(\kappa^2).
\end{aligned}
\end{equation}
The leading width, $\tilde\sigma_+^2\simeq1/\Phi''(\tilde x_+)=1/(2\kappa)$, is the local Gaussian result. The inward shift reflects the asymmetry of each effective well: its inner flank, toward the barrier, is softer than its outer flank, which displaces the population toward the trap center by $\tfrac34$, independently of $\kappa$. Carrying the expansion of $e^{-\Phi}$ to second order shows that the same asymmetry also broadens the population beyond the Gaussian estimate, $\kappa^2\tilde\sigma_+^2=\tfrac12\kappa+\tfrac{15}{8}\kappa^2+O(\kappa^3)$. In $\langle\tilde x^2\rangle_{\rm st}$, the inward shift contributes $2\kappa^{-1}\langle\xi\rangle_+=-\tfrac32\kappa^{-1}$, which outweighs the intrawell spread $+\tfrac12\kappa^{-1}$.

\textit{Strong noise, $\kappa\gg1$.}---The particles now explore $|\tilde x|\gg\kappa^{-1}$. Writing $\Phi=\tfrac14\kappa^3\tilde x^4-\tfrac12\kappa\tilde x^2+\tfrac14\kappa^{-1}$, the quartic term sets the typical scale $\tilde x\sim\kappa^{-3/4}$, on which the quadratic term is of order $\kappa^{-1/2}$ and can be neglected. Hence $e^{-\Phi}\propto e^{-\kappa^3\tilde x^4/4}$, and the integral $\int_0^\infty\tilde x^n e^{-a\tilde x^4}\,\mathrm d\tilde x=\Gamma\!\left(\frac{n+1}{4}\right)/\left(4a^{(n+1)/4}\right)$ gives
\begin{equation}
\langle\tilde x^n\rangle_+=\left(\frac{4}{\kappa^3}\right)^{n/4}\frac{\Gamma\!\left(\frac{n+1}{4}\right)}{\Gamma\!\left(\frac14\right)}.
\end{equation}
Thus $\langle\tilde x^2\rangle_{\rm st}\simeq[2\Gamma(3/4)/\Gamma(1/4)]\,\kappa^{-3/2}\approx0.676\,\kappa^{-3/2}$, and
\begin{equation}
\tilde\sigma_+^2\simeq2\left[\frac{\Gamma(3/4)}{\Gamma(1/4)}-\frac{\pi}{\Gamma(1/4)^2}\right]\kappa^{-3/2}\approx0.198\,\kappa^{-3/2},
\qquad
\kappa^2\tilde\sigma_+^2\approx0.198\sqrt{\kappa}.
\label{eq:app_strong}
\end{equation}
The corresponding force magnitude, $|\tilde f|=\kappa|\tilde x|\sim\kappa^{1/4}\gg1$, lies on the positive-mobility branch, $\tilde v\simeq\tilde f^3$. The width decreases as $\kappa^{-3/2}$, faster than the $\kappa^{-1}$ of ordinary harmonic confinement, because the cubic restoring drift $\kappa^3\tilde x^3$ strengthens with distance from the trap center.

\textit{Crossover.}---The two limits meet at $\Delta\Phi\sim1$, i.e., $\kappa\sim1/4$, where neither approximation applies and Eq.~\eqref{eq:app_sigma} must be evaluated directly. The variance relative to the squared attractor distance, $\kappa^2\langle\tilde x^2\rangle_{\rm st}$, decreases as $1-\kappa$ at weak noise but grows as $0.676\sqrt{\kappa}$ at strong noise. Evaluating Eq.~\eqref{eq:app_sigma} locates its minimum at $\kappa\simeq0.213$, where $\kappa^2\langle\tilde x^2\rangle_{\rm st}\simeq0.831$.

\section{Fokker--Planck solution of the escape problem}
\label{SI:escape}

For the cubic trap with $\kappa\tilde a=1$, the effective drift potential follows from $\Phi'=\tilde f-\tilde f^3$ with $\tilde f=(\kappa\tilde x-1)(\kappa\tilde x-2)$. Up to an additive constant,
\begin{equation}
 \Phi(\tilde x)=\frac{1}{\kappa}\left(-\frac{y^7}{7}+\frac{3y^6}{2}-\frac{33y^5}{5}+\frac{63y^4}{4}-\frac{65y^3}{3}+\frac{33y^2}{2}-6y\right),
 \qquad y=\kappa\tilde x.
 \label{eq:app_Phi_cubic}
\end{equation}
The lines in Figs.~3(b) and 3(c) of the main text are obtained from the Fokker--Planck equation $\partial_{\tilde t}\tilde p=\partial_{\tilde x}(\Phi'\tilde p)+\partial_{\tilde x}^2\tilde p$ with the initial condition $\tilde p(\tilde x,0)=\delta(\tilde x-\tilde m_1)$, the absorbing boundary $\tilde p(\tilde s_2,\tilde t)=0$, and confinement as $\tilde x\to-\infty$. The populations of the initial and intermediate basins are
\begin{equation}
 P_{\mathrm I}(\tilde t)=\int_{-\infty}^{\tilde s_1}\tilde p\,\mathrm d\tilde x,
 \qquad
 P_{\mathrm{II}}(\tilde t)=\int_{\tilde s_1}^{\tilde s_2}\tilde p\,\mathrm d\tilde x,
\end{equation}
and the survival probability $S=P_{\mathrm I}+P_{\mathrm{II}}$ satisfies $\int_0^\infty S\,\mathrm d\tilde t=\tilde\tau(\tilde m_1)$.

We discretize the Fokker--Planck equation on a uniform grid $\tilde x_i$ with spacing $\tilde h$ and define $\Phi_i=\Phi(\tilde x_i)$. The nearest-neighbor rates
\begin{equation}
 q_i^+=\frac{1}{\tilde h^2}e^{-(\Phi_{i+1}-\Phi_i)/2},
 \qquad
 q_{i+1}^-=\frac{1}{\tilde h^2}e^{+(\Phi_{i+1}-\Phi_i)/2}
\end{equation}
preserve detailed balance with respect to $\pi_i\propto e^{-\Phi_i}$ and recover the drift and unit diffusivity as $\tilde h\to0$. A reflecting end sets the outward rate to zero. An absorbing node is removed from the unknowns, so that jumps into it remove probability. For the discrete backward generator $(Qg)_i=q_i^+(g_{i+1}-g_i)+q_i^-(g_{i-1}-g_i)$, with $g=0$ at an absorbing node, the matrix $H=-\Pi^{1/2}Q\Pi^{-1/2}$, with $\Pi=\operatorname{diag}(\pi_i)$, is symmetric and tridiagonal,
\begin{equation}
 H_{ii}=q_i^++q_i^-,
 \qquad
 H_{i,i+1}=H_{i+1,i}=-\frac{1}{\tilde h^2},
\end{equation}
and its eigenproblem is solved with \texttt{eigh\_tridiagonal}.

For the escape problem, the grid extends from $\tilde x_L=-0.2/\kappa$ to $\tilde s_2$, with $\tilde h\simeq0.02$ adjusted so that $\tilde s_2$ is a grid node. The left end is reflecting, and $\tilde s_2$ is absorbing. Because of the absorption, all eigenvalues are positive,
\begin{equation}
 H\psi_n=\tilde\mu_n\psi_n,
 \qquad
 \psi_n^{\mathsf T}\psi_m=\delta_{nm},
 \qquad
 \tilde\mu_n>0,
\end{equation}
and no stationary mode exists. We retain the 60 slowest modes. The initial delta function is assigned to the two nodes bracketing $\tilde m_1$, with weights that preserve its mass and mean. Denoting by $u_i(\tilde t)$ the probability mass at node $i$, the solution $u(\tilde t)=\Pi^{1/2}e^{-H\tilde t}\Pi^{-1/2}u(0)$ reads
\begin{equation}
 u_i(\tilde t)=\sum_{n}W_{in}e^{-\tilde\mu_n\tilde t},
 \qquad
 W_{in}=e^{-\Phi_i/2}\psi_n(i)\sum_j e^{\Phi_j/2}\psi_n(j)\,u_j(0).
 \label{eq:absorbing-modes}
\end{equation}
The density in Fig.~3(b) is $\tilde p(\tilde x_i,\tilde t)=u_i(\tilde t)/\tilde h$, and $P_{\mathrm I}$ and $P_{\mathrm{II}}$ in Fig.~3(c) are the sums of $u_i$ over $\tilde x_i<\tilde s_1$ and $\tilde s_1\leq\tilde x_i<\tilde s_2$, respectively. All probabilities are normalized by the initial number of particles, as in the simulations. Times are measured in units of the exact MFPT $\tilde\tau$, with density snapshots at $\tilde t/\tilde\tau=0.002$, $0.03$, $0.15$, $0.5$, and $1.5$ and populations at 400 logarithmically spaced times in $10^{-3}\leq\tilde t/\tilde\tau\leq10$.

The solid curve in Fig.~3(d) evaluates Eq.~(7) of the main text by nested adaptive quadrature, with $\Phi$ measured from $\Phi(\tilde m_1)$ to avoid overflow. The dashed curve is $1/(\tilde\tau_{11}+\tilde\tau_{21}+\tilde\tau_{22})$ from Eq.~(9) of the main text, with $\Phi''(\tilde m_1)=-\Phi''(\tilde s_2)=2\sqrt5\,\kappa$ and $\Phi''(\tilde m_2)=-\Phi''(\tilde s_1)=\kappa$. Both are evaluated at 121 equally spaced values of $\kappa$ in $[0.018,0.052]$.

As a check, Eq.~\eqref{eq:absorbing-modes} gives $\int_0^\infty S\,\mathrm d\tilde t=\sum_n\sum_iW_{in}/\tilde\mu_n$, which reproduces Eq.~(7) of the main text at $\kappa=0.02$ to a relative accuracy of $1.5\times10^{-6}$. Halving $\tilde h$, extending the left end to $\tilde x=-0.4/\kappa$, or doubling the number of modes changes this value by less than $2\times10^{-6}$. At $\tilde x_L$, $\Phi$ exceeds $\Phi(\tilde m_1)$ by about $143$, so the reflecting end has no effect. The retained modes carry all but $10^{-4}$ of the initial mass, and the omitted modes decay faster than the fastest retained one, whose relaxation time is about $0.4$, far below the earliest plotted time $10^{-3}\tilde\tau\simeq4\times10^2$.

\section{Numerical spectral expansion}
\label{SI:Numerical spectral expansion}

In the harmonic trap, the unperturbed drift is $\tilde B_0(\tilde x)=\kappa\tilde x-\kappa^3\tilde x^3$, with stationary density $\tilde p_{\mathrm{st}}(\tilde x)=\tilde Z^{-1}e^{-\Phi(\tilde x)}$ and $\Phi$ given by Eq.~\eqref{eq:app_Phi}. The zero-current condition makes the backward generator $\tilde{\mathcal L}=\tilde B_0\partial_{\tilde x}+\partial_{\tilde x}^2$ self-adjoint in the stationary inner product $\langle hg\rangle_{\mathrm{st}}=\int_{\mathbb R}hg\,\tilde p_{\mathrm{st}}\,\mathrm d\tilde x$. Its relaxation modes satisfy
\begin{equation}
 -\tilde{\mathcal L}\varphi_n=\tilde\lambda_n\varphi_n,
 \qquad
 \langle\varphi_n\varphi_m\rangle_{\mathrm{st}}=\delta_{nm}.
\end{equation}
Expanding the position observable in these modes gives
\begin{equation}
 \tilde c_n=\langle\tilde x\varphi_n\rangle_{\mathrm{st}}^2,
 \qquad \tilde C(\tilde t)=\sum_{n\geq1}\tilde c_ne^{-\tilde\lambda_n\tilde t}.
\end{equation}
Here, $n=0$ denotes the stationary mode, and the remaining eigenvalues are ordered increasingly.

Since the mechanical force translates the drift, $\tilde B_F(\tilde x)=\tilde B_0(\tilde x-\tilde F/\kappa)$, the relation $\tilde R_{x,F}(\tilde t)=\kappa^{-1}\,\mathrm d^2\tilde C/\mathrm d\tilde t^2$ yields
\begin{equation}
 \begin{aligned}
 \rho_n&=\tilde\lambda_n\tilde c_n,\\
 \tilde R_{x,F}(\tilde t)&=\frac{1}{\kappa}\sum_{n\geq1}\rho_n\tilde\lambda_ne^{-\tilde\lambda_n\tilde t},\\
 \tilde\chi(\tilde t)&=\frac{1}{\kappa}\sum_{n\geq1}\rho_n\left(1-e^{-\tilde\lambda_n\tilde t}\right).
 \end{aligned}
\end{equation}
For the passive unit-mobility particle in the same $\Phi$, the additive force gives $\tilde\chi^{\mathrm p}(\tilde t)=\tilde C(0)-\tilde C(\tilde t)$, that is,
\begin{equation}
 \rho_n^{\mathrm p}=\frac{\tilde c_n}{\langle\tilde x^2\rangle_{\mathrm{st}}},
 \qquad
 \tilde\chi^{\mathrm p}(\tilde t)=\langle\tilde x^2\rangle_{\mathrm{st}}\sum_{n\geq1}\rho_n^{\mathrm p}\left(1-e^{-\tilde\lambda_n\tilde t}\right).
\end{equation}
Moreover, $\sum_n\tilde\lambda_n\tilde c_n=-\langle\tilde x\,\tilde{\mathcal L}\tilde x\rangle_{\mathrm{st}}=1$, so $\sum_n\rho_n=1$ and $\tilde\chi(\infty)=1/\kappa$, whereas $\sum_n\rho_n^{\mathrm p}=1$ and $\tilde\chi^{\mathrm p}(\infty)=\langle\tilde x^2\rangle_{\mathrm{st}}$. Thus, $\tilde c_n$ measures position memory, whereas $\rho_n$ and $\rho_n^{\mathrm p}$ measure the contribution of mode $n$ to the step response of the two particles.

The density changes in Figs.~4(d) and 4(g) of the main text follow from the same modes. After the force is switched on, the density relaxes from $\tilde p_{\mathrm{st}}$ toward the new stationary density, whose linear change is known exactly. We write it as $\tilde F\tilde G\,\tilde p_{\mathrm{st}}$, with $\tilde G=-\tilde B_0/\kappa$ for the translated potential of the negative-mobility particle and $\tilde G=\tilde x$ for the tilted potential of the passive particle. To linear order, the deviation from the new stationary density obeys the unperturbed Fokker--Planck equation, so each mode component relaxes at its own rate:
\begin{equation}
 \frac{\delta\tilde p(\tilde x,\tilde t)}{\tilde F}
 =\tilde p_{\mathrm{st}}(\tilde x)\sum_{n\geq1}\langle\tilde G\varphi_n\rangle_{\mathrm{st}}\left(1-e^{-\tilde\lambda_n\tilde t}\right)\varphi_n(\tilde x),
 \label{eq:density-modes}
\end{equation}
and likewise for $\delta\tilde p^{\mathrm p}$. Since $\tilde B_0=\tilde{\mathcal L}\tilde x$, the projections are $\langle\tilde G\varphi_n\rangle_{\mathrm{st}}=\tilde\lambda_n\langle\tilde x\varphi_n\rangle_{\mathrm{st}}/\kappa$ for the negative-mobility particle and $\langle\tilde x\varphi_n\rangle_{\mathrm{st}}$ for the passive particle. Multiplying Eq.~\eqref{eq:density-modes} by $\tilde x$ and integrating therefore recovers $\tilde\chi$ and $\tilde\chi^{\mathrm p}$, respectively. These results assume stationary initial conditions, unit diffusion coefficient, and vanishing boundary terms on the full line.

To compute the coefficients, we use the discretization of Sec.~\ref{SI:escape} on the grid $\tilde x_i=-\tilde L+i\tilde h$, with reflecting ends at $\pm\tilde L$ and normalized stationary masses $\pi_i=e^{-\Phi_i}/\sum_j e^{-\Phi_j}$. Without absorption, $H$ has the stationary eigenvector $\psi_0(i)=\sqrt{\pi_i}$ with $\tilde\lambda_0=0$. Its eigenvectors, $H\psi_n=\tilde\lambda_n\psi_n$ with $\psi_n^{\mathsf T}\psi_m=\delta_{nm}$, give the modes on the grid as $\varphi_n(\tilde x_i)=\psi_n(i)/\sqrt{\pi_i}$, and we compute
\begin{equation}
 \tilde c_n\simeq
 \left(\sum_i \tilde x_i\sqrt{\pi_i}\,\psi_n(i)\right)^2,
 \qquad \rho_n=\tilde\lambda_n\tilde c_n,
 \qquad \rho_n^{\mathrm p}=\frac{\tilde c_n}{\sum_i\pi_i\tilde x_i^2}.
\end{equation}
For the density, the projections and the mode profiles become
\begin{equation}
 \langle\tilde G\varphi_n\rangle_{\mathrm{st}}\simeq\sum_i\tilde G(\tilde x_i)\sqrt{\pi_i}\,\psi_n(i),
 \qquad
 \tilde p_{\mathrm{st}}(\tilde x_i)\varphi_n(\tilde x_i)\simeq\frac{\sqrt{\pi_i}\,\psi_n(i)}{\tilde h},
\end{equation}
which are inserted into Eq.~\eqref{eq:density-modes}. No additional factor of $\tilde h$ is required in the sums because $\pi_i$ is a probability mass. We use the spacing $\tilde h=5\times10^{-3}$, which resolves the intrawell width $(2\kappa)^{-1/2}\geq1.58$ with more than 300 grid points. The domain extends to $\tilde L=2.5\,\tilde x_+=2.5/\kappa$, where $\Phi(\tilde L)-\Phi(\tilde x_+)\approx6.9/\kappa\geq34$ for the values considered, so truncation effects are negligible. The response weights retain 20 modes, including the stationary mode, without renormalization. The density changes retain 60 modes and are evaluated at $\kappa\tilde t=0.5$, $2$, $20$, and $360$.

\begin{table*}[htbp]
 \centering
 \caption{The first five nonstationary eigenmodes and the response weights of the negative-mobility particle, in the dimensionless units of the main text. The column $\tilde\lambda_n/\kappa$ compares each relaxation rate with the trap rate; the intrawell rate is $\Phi''(\tilde x_+)/\kappa=2$. Even modes have $\tilde c_n=\rho_n=0$ by symmetry.}
 \label{tab:linear-response-spectrum}
\begin{tabular}{c@{\hspace{0.8cm}}crrrr}
 \hline
 $\kappa$ & $n$ & $\tilde\lambda_n$ & $\tilde\lambda_n/\kappa$ & $\tilde c_n$ & $\rho_n$\\
 \hline
 0.05 & 1 & $1.38807\times10^{-4}$ & 0.00277614 & 363.147 & 0.0504073\\
      & 2 & 0.0411124 & 0.822247 & 0 & 0\\
      & 3 & 0.0657078 & 1.31416 & 8.81425 & 0.579165\\
      & 4 & 0.0890460 & 1.78092 & 0 & 0\\
      & 5 & 0.122752 & 2.45505 & 2.46237 & 0.302261\\
 \hline
 0.10 & 1 & 0.00335453 & 0.0335453 & 83.4039 & 0.279781\\
      & 2 & 0.0927372 & 0.927372 & 0 & 0\\
      & 3 & 0.168026 & 1.68026 & 3.32216 & 0.558210\\
      & 4 & 0.259582 & 2.59582 & 0 & 0\\
      & 5 & 0.373398 & 3.73398 & 0.380616 & 0.142121\\
 \hline
 0.20 & 1 & 0.0269556 & 0.134778 & 19.9359 & 0.537384\\
      & 2 & 0.262516 & 1.31258 & 0 & 0\\
      & 3 & 0.506386 & 2.53193 & 0.781310 & 0.395644\\
      & 4 & 0.808234 & 4.04117 & 0 & 0\\
      & 5 & 1.16658 & 5.83289 & 0.0522850 & 0.0609946\\
 \hline
 \end{tabular}
\end{table*}

\begin{table*}[htbp]
 \centering
 \caption{Same as Table~\ref{tab:linear-response-spectrum}, for the passive unit-mobility particle in the same effective drift potential $\Phi$. The relaxation rates and position weights coincide with those of the negative-mobility particle, whereas the response weights are $\rho_n^{\mathrm p}$, with $\langle\tilde x^2\rangle_{\mathrm{st}}=374.734$, $87.1363$, and $20.7724$ for $\kappa=0.05$, $0.10$, and $0.20$. Even modes have $\tilde c_n=\rho_n^{\mathrm p}=0$ by symmetry.}
 \label{tab:passive-spectrum}
\begin{tabular}{c@{\hspace{0.8cm}}crrrr}
 \hline
 $\kappa$ & $n$ & $\tilde\lambda_n$ & $\tilde\lambda_n/\kappa$ & $\tilde c_n$ & $\rho_n^{\mathrm p}$\\
 \hline
 0.05 & 1 & $1.38807\times10^{-4}$ & 0.00277614 & 363.147 & 0.969079\\
      & 2 & 0.0411124 & 0.822247 & 0 & 0\\
      & 3 & 0.0657078 & 1.31416 & 8.81425 & 0.0235214\\
      & 4 & 0.0890460 & 1.78092 & 0 & 0\\
      & 5 & 0.122752 & 2.45505 & 2.46237 & 0.00657098\\
 \hline
 0.10 & 1 & 0.00335453 & 0.0335453 & 83.4039 & 0.957166\\
      & 2 & 0.0927372 & 0.927372 & 0 & 0\\
      & 3 & 0.168026 & 1.68026 & 3.32216 & 0.0381260\\
      & 4 & 0.259582 & 2.59582 & 0 & 0\\
      & 5 & 0.373398 & 3.73398 & 0.380616 & 0.00436805\\
 \hline
 0.20 & 1 & 0.0269556 & 0.134778 & 19.9359 & 0.959731\\
      & 2 & 0.262516 & 1.31258 & 0 & 0\\
      & 3 & 0.506386 & 2.53193 & 0.781310 & 0.0376129\\
      & 4 & 0.808234 & 4.04117 & 0 & 0\\
      & 5 & 1.16658 & 5.83289 & 0.0522850 & 0.00251705\\
 \hline
 \end{tabular}
\end{table*}
Halving $\tilde h$ changes $\tilde\lambda_n$, $\tilde c_n$, and $\rho_n$ for the first five contributing modes ($n=1,3,5,7,9$) by less than $4.1\times10^{-5}$ relatively. The retained sums of $\rho_n$ are 0.99999972, 0.99999950, and 0.99999907 for $\kappa=0.05$, $0.10$, and $0.20$, respectively. For the density changes, the $\tilde t\to\infty$ limit of Eq.~\eqref{eq:density-modes} reproduces the exact stationary change $\tilde F\tilde G\,\tilde p_{\mathrm{st}}$ to a relative accuracy of $10^{-8}$. At $\kappa=0.05$, the slowest mode has the relaxation time $1/\tilde\lambda_1\simeq7.20\times10^3$, corresponding to $\kappa/\tilde\lambda_1\simeq360$ in the units $\kappa\tilde t$ of Fig.~4 of the main text. It carries about $97\%$ of $\langle\tilde x^2\rangle_{\rm st}$ but only $5.04\%$ of the step response of the negative-mobility particle, compared with $96.9\%$ for the passive particle. The contributing faster modes relax at $\tilde\lambda_3\simeq1.31\kappa$ and $\tilde\lambda_5\simeq2.46\kappa$, comparable to the intrawell rate $2\kappa$. The factor $\tilde\lambda_n$ in $\rho_n=\tilde\lambda_n\tilde c_n$ thus suppresses the response weight of long-lived position memory.